\documentclass[apj,twocolumn]{emulateapj}
\usepackage{graphicx, color}
\usepackage{afterpage}
\usepackage{amssymb}
\def\tabmultiples {
\tabletypesize{\tiny}
\setlength{\tabcolsep}{0.03in}
\begin{deluxetable*}{|c|c|c|c|c|c|c|c|}
\tablecolumns{8}
\tablewidth{0pt}
\tablecaption{\small{Multiple Images and Candidates} \label{multTable}}
\tablehead{
\colhead{Arc ID
} &
\colhead{$\alpha$(deg.)
} &
\colhead{$\delta$(deg.)
} &
\colhead{Phot-$z$ [95\% C.I.]
} &
\colhead{Input $z$
} &
\colhead{$z_{LTM}$ [95\% C.I.]
} &
\colhead{$z_{NFW}$ [95\% C.I.]
} &
\colhead{Comments
}
}
\startdata
A209   &                &                 &       &      &      &    &  \\
1.1 &  22.966446 & -13.611836 & 1.13 [1.08,1.23] & 2.63 &      \nodata             & \nodata  &            \\
1.2 &  22.967486 & -13.610139 & \nodata              & "    &      \nodata             & \nodata &   \\
1.3 &  22.970421 & -13.608081 & 2.63 [2.49,2.73] & "    &      \nodata              & \nodata&   \\
2.1 &  22.968694 & -13.615574 & 0.64 [0.55,0.69] & \nodata  &  3.47 [2.97,5.83] & 0.77 [0.67,0.85] & \\
2.2 &  22.967876 & -13.616512 & \nodata              & \nodata  &      "             & " &   \\
3.1 &  22.967951 & -13.615285 & \nodata              & \nodata  &  1.74 [1.74,4.90]  & 4.78 [2.14,6.21] & \\
3.2 &  22.968025 & -13.616946 & \nodata              & \nodata  &      "             & " &  \\
\enddata
\end{deluxetable*}

    } 

\begin{document}
\shorttitle{HST SL+WL Analysis of the CLASH sample}
\shortauthors{Zitrin \& the CLASH collaboration}

\slugcomment{Submitted to the Astrophysical Journal}

\title{Hubble Space Telescope Combined Strong and Weak Lensing Analysis of the CLASH Sample: Mass and Magnification Models and Systematic Uncertainties}


\author{Adi Zitrin\altaffilmark{1,2}, Agnese Fabris\altaffilmark{3}, Julian Merten\altaffilmark{4,1}, Peter Melchior\altaffilmark{5}, Massimo Meneghetti\altaffilmark{4}, Anton Koekemoer\altaffilmark{6}, Dan Coe\altaffilmark{6}, Matteo Maturi\altaffilmark{3}, Matthias Bartelmann\altaffilmark{3}, Marc Postman\altaffilmark{6}, Keiichi Umetsu\altaffilmark{7}, Gregor Seidel\altaffilmark{8}, Irene Sendra\altaffilmark{9}, Tom Broadhurst\altaffilmark{9,10}, Italo Balestra\altaffilmark{11}, Andrea Biviano\altaffilmark{11}, Claudio Grillo\altaffilmark{12}, Amata Mercurio\altaffilmark{13}, Mario Nonino\altaffilmark{11}, Piero Rosati\altaffilmark{14}, Larry Bradley\altaffilmark{6}, Mauricio Carrasco\altaffilmark{3}, Megan Donahue\altaffilmark{15}, Holland Ford\altaffilmark{16}, Brenda L. Frye\altaffilmark{17}, John Moustakas\altaffilmark{18}}

\altaffiltext{1}{Cahill Center for Astronomy and Astrophysics, California Institute of Technology, MC 249-17, Pasadena, CA 91125, USA; adizitrin@gmail.com}
\altaffiltext{2}{Hubble Fellow}
\altaffiltext{3}{Universit\"at Heidelberg, Zentrum f\"ur Astronomie, Institut f\"ur Theoretische Astrophysik, Philosophenweg 12, 69120 Heidelberg, Germany}
\altaffiltext{4}{Jet Propulsion Laboratory, California Institute of Technology, Pasadena, CA 91109, USA}
\altaffiltext{5}{Center for Cosmology and Astro-Particle Physics \& Department of Physics, The Ohio State University, Columbus, OH, 43210, USA}
\altaffiltext{6}{Space Telescope Science Institute, 3700 San Martin Drive, Baltimore MD 21218, USA}
\altaffiltext{7}{Institute of Astronomy and Astrophysics, Academia Sinica, PO Box 23-141, Taipei 10617, Taiwan}
\altaffiltext{8}{Max-Planck-Institut f\"ur Astronomie, K\"onigstuhl 17, D-69117 Heidelberg, Germany}
\altaffiltext{9}{Department of Theoretical Physics, University of Basque Country UPV/EHU, Bilbao, Spain}
\altaffiltext{10}{IKERBASQUE, Basque Foundation for Science, Bilbao, Spain}
\altaffiltext{11}{INAF/Osservatorio Astronomico di Trieste, via G.B. Tiepolo 11, I-34143 Trieste, Italy}
\altaffiltext{12}{Dark Cosmology Centre, Niels Bohr Institute, University of Copenhagen, Juliane Maries Vej 30, DK-2100 Copenhagen, Denmark}
\altaffiltext{13}{INAF/Osservatorio Astronomico di Capodimonte, via Moiariello 16, I-80131 Napoli, Italy}
\altaffiltext{14}{Department of Physics and Earth Science, University of Ferrara, Via G.Saragat, 1-44122 Ferrara, Italy}
\altaffiltext{15}{Department of Physics and Astronomy, Michigan State University, East Lansing, MI 48824, USA}
\altaffiltext{16}{Department of Physics and Astronomy, Johns Hopkins University, Baltimore, MD 21218, USA}
\altaffiltext{17}{Steward Observatory, Department of Astronomy, University of Arizona, 933 N. Cherry Avenue, Tucson, AZ 85721, USA}
\altaffiltext{18}{Department of Physics \& Astronomy, Siena College, 515 Loudon Road, Loudonville, NY 12211, USA}

\begin{abstract}
We present results from a comprehensive lensing analysis in HST data, of the complete Cluster Lensing And Supernova survey with Hubble (CLASH) cluster sample. We identify new multiple-images previously undiscovered, allowing improved or first constraints on the cluster inner mass distributions and profiles. We combine these strong-lensing constraints with weak-lensing shape measurements within the HST FOV to jointly constrain the mass distributions. The analysis is performed in two different common parameterizations (one adopts light-traces-mass for both galaxies and dark matter while the other adopts an analytical, elliptical NFW form for the dark matter), to provide a better assessment of the underlying systematics - which is most important for deep, cluster-lensing surveys, especially when studying  magnified high-redshift objects. We find that the typical (median), relative systematic differences throughout the central FOV are $\sim40\%$ in the (dimensionless) mass density, $\kappa$, and $\sim20\%$ in the magnification, $\mu$. We show maps of these differences for each cluster, as well as the mass distributions, critical curves, and 2D integrated mass profiles. For the Einstein radii ($z_{s}=2$) we find that all typically agree within $10\%$ between the two models, and Einstein masses agree, typically, within $\sim15\%$. At larger radii, the total projected, 2D integrated mass profiles of the two models, within $r\sim2\arcmin$, differ by $\sim30\%$. Stacking the surface-density profiles of the sample from the two methods together, we obtain an average slope of $d\log (\Sigma)/d\log(r)\sim-0.64\pm0.1$, in the radial range [5,350] kpc. Lastly, we also characterize the behavior of the average magnification, surface density, and shear differences between the two models, as a function of both the radius from the center, and the best-fit values of these quantities. All mass models and magnification maps are made publicly available for the community.
\end{abstract}



\section{Introduction}\label{intro}

Lensing by galaxy clusters has become of great interest, due to the inherent ability to constrain the underlying matter distribution of the lens, dominated by an \emph{unseen} dark matter (DM) component, and thanks to the magnification effect that distorts and enhances faint background objects to be detected through such cosmic lenses \citep[e.g.][for recent reviews]{Bartelmann2010reviewB,Kneib2011review}.

Background galaxies lensed by galaxy clusters are magnified in size and flux, and get distorted and sheared, as a consequence of the cluster's gravitational potential. In recent years, the inner parts of galaxy clusters have been mapped with increasing precision, particularly through the strong-lensing (SL) phenomenon in which background galaxies are also multiply-imaged, allowing for high-resolution constraints to be placed on the mass distribution and profile \citep[][as a few examples]{Kneib2004z7,Broadhurst2005a,Smith2005,Limousin2008,Newman2009,Richard2010locuss20,Richard2014FF, Bradac2008rxj1347,Liesenborgs2008CL0024L,Diego2005Nonparam,Diego2014M0416,Diego2014A1689,Coe2010,Oguri2012SL,Sereno2013A1689D3,Zitrin2009b,Zitrin2011_12macsclusters,Zitrin2013Gordo,Jauzac2014M0416,Grillo2014_0416}. In particular, this improvement is attributed to the remarkable spatial resolution and image quality of the Hubble Space Telescope (HST) that has allowed the detection of many multiple-image constraints in clusters, as reflected in the works mentioned above. This has become well-acknowledged, motivating substantial cluster lensing surveys, such as the Cluster Lensing and Supernova survey with Hubble (CLASH; \citealt{PostmanCLASHoverview}) in which 25 clusters were observed in 16 filters so that many multiple images could be found, and their redshifts well determined, allowing to map the cluster mass distributions with great precision; or the Hubble Frontier Fields program (HFF)\footnotemark[1] \footnotetext[1]{http://www.stsci.edu/hst/campaigns/frontier-fields/} set to observe 4-6 massive clusters to an unprecedented depth, to exploit their magnification power (and our ability to map it through lensing) and study very high-$z$ galaxies (\citealt{Atek2014A2744,Atek2014LF2,Coe2014FF,Ishigaki2014,Laporte2014FF,Zheng2014A2744,Zitrin2014highz}, see also \citealt{Lotz2014AAS_FFreview}).

Further outwards from the cluster core, where the density is typically lower than the critical density for strong lensing \citep[see][]{NarayanBartelmann1996Lectures}, background objects observed through the lensing cluster will be (only) slightly sheared and magnified, an effect that could be detected, in principle, only on a statistical basis due to the intrinsic scatter in their source ellipticities \citep[][]{KSB1995,BartelmannShneider2001WLreview,HoekstraWL2008}. This weak-lensing (WL) effect is thus used to map the mass distribution (or profile) out to the virial radius and beyond, allowing a large-scale view of the cluster and surrounding structures  \citep[e.g.][]{Merten2009,Umetsu2009,Umetsu2010,Umetsu2012,Okabe2010_30WL,Jauzac2012WL0717,Medezinski2013M0717}.

Among the main goals of the CLASH program is addressing some standing questions related to structure formation in the context of the standard $\Lambda$CDM paradigm. Accurate mass maps for the clusters can be exploited for characterizing with unprecedented precision the observational concentration-mass relation, and Einstein radius distribution, for example, both of which have been previously claimed to be in some tension with predictions from semi-analytic calculations or simulations based on $\Lambda$CDM \citep[e.g.][]{BroadhurstBarkana2008,Broadhurst2008,Zitrin2011_12macsclusters,Zitrin2010_A1703,Meneghetti2011}.

Various studies have previously combined SL+WL \citep[e.g.][]{Bradac2006Bullet,LimousinOnA1689,Merten2009,Merten2011,Umetsu2011,Umetsu2011b,Umetsu2012,Oguri201238clusters,Newman2013,Newman2013b}. These, however, are often made either (a) independently (i.e. after the fact, so that each regime is first used to construct a mass model, regardless of the other regime's constraints), (b) non-parametrically, meaning, without any assumptions on the mass distribution or use of a parameterized model, but using a (usually lower-resolution) free-form grid instead, or (c) using wide-field ground based imaging for the WL regime. Here, we aim to combine the two effects for a simultaneous fit in HST data alone \citep[e.g.][]{Kneib1996,Smith2005,Richard2014FF}, through a joint minimization of a high-resolution parametric model. Although the HST FOV is smaller than typical wide-field imaging, its remarkable resolution allows for shape measurements of a higher number density of background galaxies \citep[e.g.][]{Kneib1996,Merten2014CLASHcM}. Moreover, we perform the fit with two distinct parameterizations, and so quantify and characterize the underlying systematic differences between them \footnote{note that throughout we may refer to these differences simply as ``systematics'' or ``systematic uncertainties'', where the meaning remains the systematic differences between these two specific methods (but can be regarded more generally as a case study of systematic uncertainties in lens modeling).}. This quantification, especially on a substantial sample, is a great leap forward in estimating the true errors on lens modeling, and is most important in the era of precision cosmology and designated deep cluster surveys aiming to study the magnified high-$z$ Universe through cluster lenses, such as the HFF.

We jointly analyze the SL and WL signals in the central HST field-of-view (FOV) of the complete sample of 25 galaxy clusters observed recently in the CLASH program. All mass models presented in this work are being made publicly available to the astronomical community through the Mikulski Archive for Space Telescopes (MAST), as a CLASH high-end science product. The models we release include \emph{*.fits} file scalable maps of the deflection fields, projected mass density, magnification, and shear components, as well as their error maps. In addition, the multiple image identification or catalogs we list here (Table \ref{multTable}) can be used for future independent modeling in other techniques, to compare to our current findings. In subsequent works (Umetsu et al., Meneghetti et al., in preparation), we aim to use the models presented here to compare the overall statistical properties of the sample (such as concentration-mass relation, or the Einstein radius distribution), with predictions by $\Lambda$CDM. In fact, as part of our broad effort to characterize structure formation CLASH has recently published the most up-to-date concentration-mass relation from wide-field SL+WL non-parametric joint analysis, while comparing it to numerical simulations \citep{Merten2014CLASHcM,Meneghetti2014CLASHsim}, and published independent WL \citep{Umetsu2014CLASH_WL} and X-ray \citep{Donahue2014CLASHX} analyses for the majority of the sample. Our following analysis concentrates on high-resolution mass and magnification mapping of the cluster cores, for the full CLASH sample.

The paper is organized as follows: In \S \ref{obs} we summarize the observations and data reduction including shape measurements. In \S \ref{Modeling} we outline the lens modeling techniques we use here, and their application to the CLASH clusters. In \S \ref{results} we briefly summarize the analysis per cluster, where the full sample modeling results are presented and discussed in \S \ref{discussion} along with the revealed systematic uncertainties or differences between the two medeling methods. The work is summarized and concluded in \S \ref{summary}. Throughout the work we use standard $\Lambda$CDM cosmology with ($\Omega_{\rm m0}=0.3$, $\Omega_{\Lambda 0}=0.7$, $H_{0}=100$ $h$ km s$^{-1}$Mpc$^{-1}$, with $h=0.7$). We often also abbreviate \emph{Abell} clusters (e.g. \citealt{Abell1989cat}) with ``A'', and MACS clusters (MAssive Cluster Survey; e.g. \citealt{EbelingMacsCat2001,Ebeling2010FinalMACS}) with ``M'', \emph{et cetera}.

\begin{figure*}
\centering
\includegraphics[width=160mm]{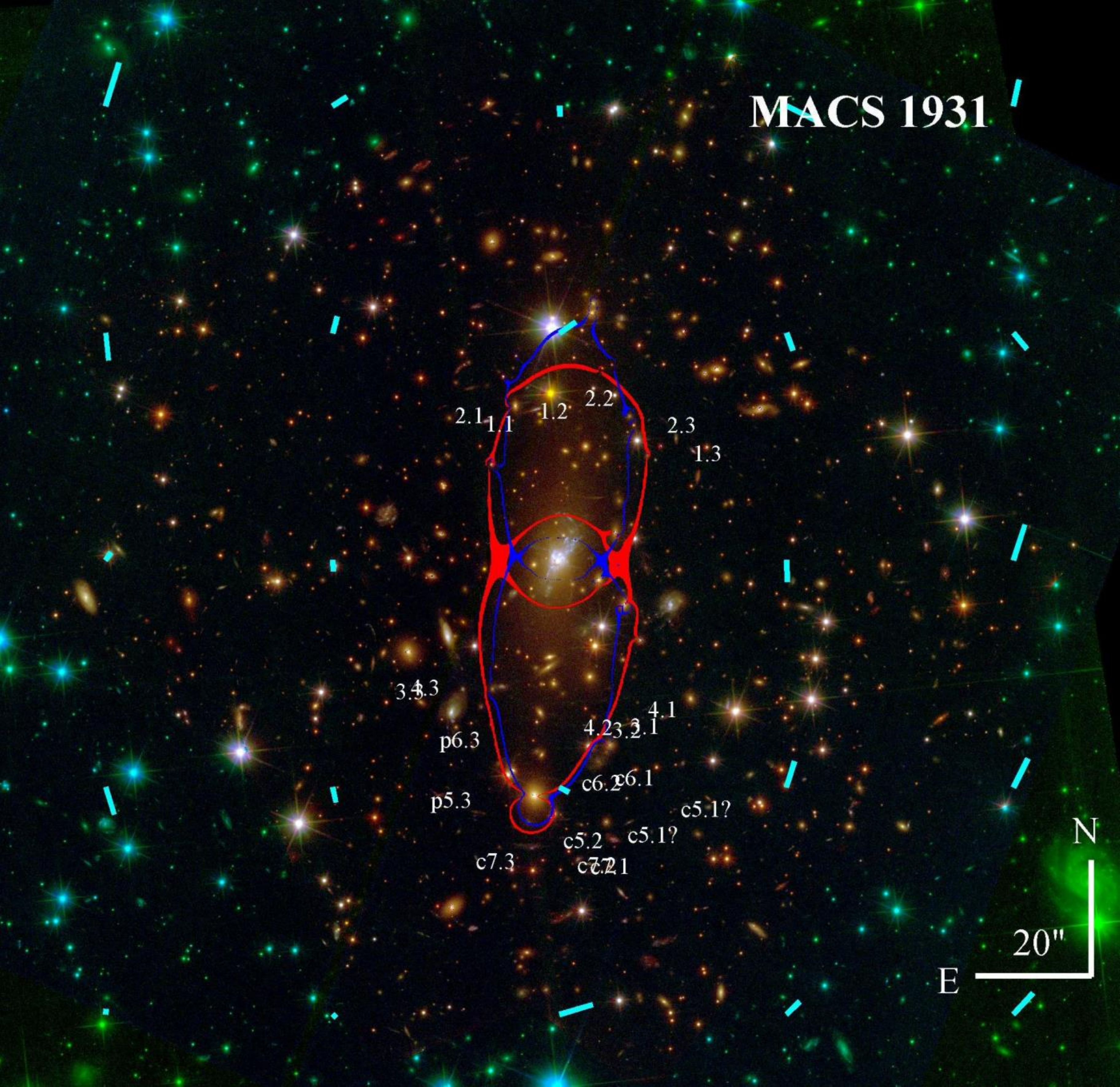}
\caption{Multiple images and candidates (the latter are marked with ``c''; ``p'' stands for predicted location), shear, and critical curves ($z_{s}=2$), overlaid on an RGB color image constructed from the CLASH 16-band imaging, for one cluster from our sample (MACS 1931). Similar Figures for the remaining 24 CLASH clusters are shown in Figs. \ref{curves1}-\ref{curves4} placed at the end of this manuscripts. The \emph{red} critical curves correspond to our LTM model whereas the \emph{blue} critical curves correspond to our PIEMDeNFW model. The measured shear, averaged here for show in $\sim[40\arcsec\times40\arcsec]$ pixels, is marked with \emph{cyan} lines across the field, where the line length in each position is proportional to the shear's strength (with the overall scale factor arbitrary). Multiple images are listed in Table \ref{multTable}; the resulting mass profiles for this cluster are shown in Fig. \ref{prof0}; the resulting mass density maps are shown in Fig. \ref{KapLTMNFW0}; and the differences between the various maps from the two models are shown in Fig. \ref{DifsKap0}.}\label{curves0}
\end{figure*}

\section{Data and Observations}\label{obs}
Each of the 25 CLASH clusters was observed with HST, generally in 16 filters ranging from the UVIS, through the optical into the near-IR, using the WFC3 and ACS cameras. Each cluster was observed to a depth of $\sim15-20$ orbits during HST's cycles 18, 19, or 20, often supplementing some existing observations. For full details, we refer the reader to \citet{PostmanCLASHoverview}. Cluster redshifts are mainly taken from \citet{PostmanCLASHoverview}, and references therein, although slight (and negligible for our purposes) discrepancies may apply e.g. due to different round-ups, revision with new spectroscopic data, or later literature.

The SL constraints, namely the positions of multiple images and their redshift information, were adopted from previous works where available and are often complemented here with newly uncovered sets. New multiple images were generally uncovered here with the aid of a preliminary model constructed for each cluster using the \citet{Zitrin2009b} method with the assumption that light-traces-mass (LTM hereafter; see \S \ref{LTM}), so that these multiple images are not simply chosen by eye but are also predicted physically by a preliminary light-tracing model. For some of these we present first spectroscopic measurements taken by the CLASH-VLT campaign (PI: Rosati). For the multiple images that lack spectroscopic data to date, we typically adopt the multi-band photometric redshift from the CLASH pipeline that incorporates the Bayesian Photometric Redshift (BPZ) software (\citealt{Benitez2000,Coe2006}; see also \citealt{Jouvel2014photoz} for CLASH photo-$z$ accuracy), although often some of these redshifts we leave as free parameters to be optimized in the minimization procedure, as we shall specify for each cluster separately (e.g. Tables \ref{ResultsTable} and \ref{multTable}). In \S \ref{results} we also include additional brief background such as previous analyses, multiple images, the ellipticity catalog, or other notable features stemming from our current analysis, upon relevancy.

For the HST WL shape measurements, we produced images with 0.03\arcsec/pixel by drizzling each visit in the unrotated frame of the ACS detector, using a modified version of the ``Mosaicdrizzle" pipeline (described more fully in \citealt{Koekemoer2011}). This allows accurate PSF treatment that does not compromise the intrinsic shape measurements required by WL pipelines. The RRG \citep{Rhodes2000WLpipe} WL shape measurement package was then used to measure shapes in each of six ACS bands (F435W, F475W, F625W, F775W, F814W, and F850LP). The RRG pipeline corrects for the Hubble PSF by determining the telescope's focus offset from the nominal value. The focus offset is determined by the inspection of stellar ellipticities in the full field and by cross-checking with the STScI focus tool\footnotemark[2] \footnotetext[2]{http://www.stsci.edu/hst/observatory/focus} for each visit's image. From the focus offset, a PSF model is created based on \citet{Rhodes2007} and shape measurements are corrected accordingly \citep[see][for more details]{Rhodes2000WLpipe,Rhodes2007,Merten2014CLASHcM}.

We exclude objects with S/N $<10$ and size $<0.1\arcsec$ since faint or poorly resolved galaxies are known to yield very inaccurate shape measurements. All the shape catalogs were then matched to the deep multi-band photometric catalog and, for objects that were successfully measured in more than a single filter, the ellipticities were combined by a S/N-weighted average to reduce the measurement noise. A selection for lensed background galaxies is achieved by choosing galaxies with a minimum photometric redshift estimate $\min(z_b) = z_c + 0.2$, such that the cluster redshift $z_c$ is well below the 95\% confidence region of the BPZ redshift distribution. Due to the faintness of the objects, no BPZ quality cuts were applied.

In \S \ref{Modeling} we now describe the lens modeling pipeline.


\section{Lens Modeling}\label{Modeling}

For the combined SL+WL analysis we use a revised version of the \citet{Zitrin2009b,Zitrin2013M0416,Zitrin2013Gordo} SL modeling technique, extended here to include also WL shape measurements, for joint minimization throughout the HST/ACS frame. The lens modeling code includes two different parameterizations, which we use here to examine the credibility of the resulting mass and magnification models and assess the underlying systematic uncertainties or differences between them. We give here a brief review of these techniques, including the extension to the WL regime, but refer the reader to the above works for further details were these required.

\subsection{Light-Traces-Mass}\label{LTM}
The first method we use here adopts the assumption that the mass distribution, of both the galaxies \emph{and} DM, is reasonably traced by the cluster's light distribution \citep{Broadhurst2005a}. The first component of the mass model is the superposition of all cluster galaxies, each modeled by a power-law surface mass density profile, scaled by its luminosity. The exponent of this power-law, $q$, is the same for all galaxies yet is a free parameter in the minimization, and thus is iterated for in each cluster. The resulting galaxies' mass map is then smoothed, using a 2D Gaussian, to represent the smoother, DM component (there is an option to use a 2D Spline interpolation smoothing instead). The Gaussian width (or polynomial degree) $S$ is the second free parameter of the model. The two components are then simply added with a relative weight, $k_{gal}$, which is also left free to be optimized by the minimization procedure. To allow for further flexibility, a 2-component external shear is then added. The amplitude and direction of the external shear are two additional free parameters. The overall normalization of the model, $K_{norm}$, is the final, free fundamental parameter. The modeling thus includes only six free fundamental parameters. The minimal number of parameters, but more so, the reasonable assumption that light traces mass, \emph{readily} allows for the detection of multiple-image sets \citep[][as few examples]{Broadhurst2005a,Zitrin2009b,Zitrin2009a,Zitrin2013Gordo,Zitrin2013M0416}.

In addition, to allow for further flexibility and since not all galaxies are expected to have the exact same mass-to-light ratio, one can allow the weight of chosen galaxies to be optimized in the minimization. Also, ellipticity, and independently, a core, can be added to specified galaxies. In fact, as a rule of thumb, we generally make use here of this feature and assign to the BCG its measured ellipticity value from SExtractor.

The best-fitting model, parameter values and errors, are obtained by a dozen to several-dozen thousand Monte-Carlo Markov Chain (MCMC) steps. The goodness-of-fit criteria for the SL regime is embedded in the form of a $\chi^2$ of the position of multiple images:

\begin{equation}\label{chi2SL}
\chi_{SL}^2=\sum_{i} \frac{(x_{i}'-x_{i})^2+(y_{i}'-y_{i})^2}{\sigma_{pos}^2},
\end{equation}

where $[x_{i},y_{i}]$ is the position of the $i$'th multiple image; $[x_{i}',y_{i}']$ is the position of the $i$'th multiple image predicted by the model; and we take throughout a positional uncertainty of $\sigma_{pos}=0.5\arcsec$ \citep[see e.g.][]{Newman2013}.

\subsection{PIEMD + eNFW}\label{PIEMDeNFW}
The second method we use here adopts the light-traces-mass assumption only for the galaxy component, whereas the DM component is obtained by adopting a symmetric, analytic form. Here, cluster galaxies are modeled each as a Pseudo-Isothermal Elliptical Mass Distribution (PIEMD) scaled by its luminosity (although note that as for the LTM model, typically we do not incorporate ellipticities for the cluster galaxies aside for the BCGs). The PIEMD prescription adopted follows \citet[][see also \citealt{Zitrin2013M0416,Zitrin2013Gordo}]{Jullo2007Lenstool}. The DM component is modeled as an elliptical NFW \citep{Navarro1996} mass-density distribution (eNFW hereafter). The velocity dispersion, $\sigma_{*}$, and the cut-off radius $r_{cut,*}$, of a reference galaxy $M_{*}$, are the two free parameters of the galaxies component. The DM component comprises four more fundamental free parameters: the mass and concentration, $M_{200}$ and $C_{200}$, the ellipticity and position angle, $e$ and $PA$, where two additional parameters, namely the 2D shift of the DM halo center from the BCG, can be added - although we usually do not make use of this feature and force the DM center to coincide with that of the BCG. Additionally, in complex, or merging clusters, it is often required to add additional DM (i.e. eNFW here) halos to well reproduce the mass distribution \citep[e.g.][]{Smith2009M1149,Limousin2012_M0717,Zitrin2013Gordo,Zitrin2013M0416}, where in our work here we limit the number of DM halos to two. For each cluster we specify in \S \ref{results} additional details relevant for its specific analysis.

\begin{figure}\hspace{-1cm}
\includegraphics[width=100mm]{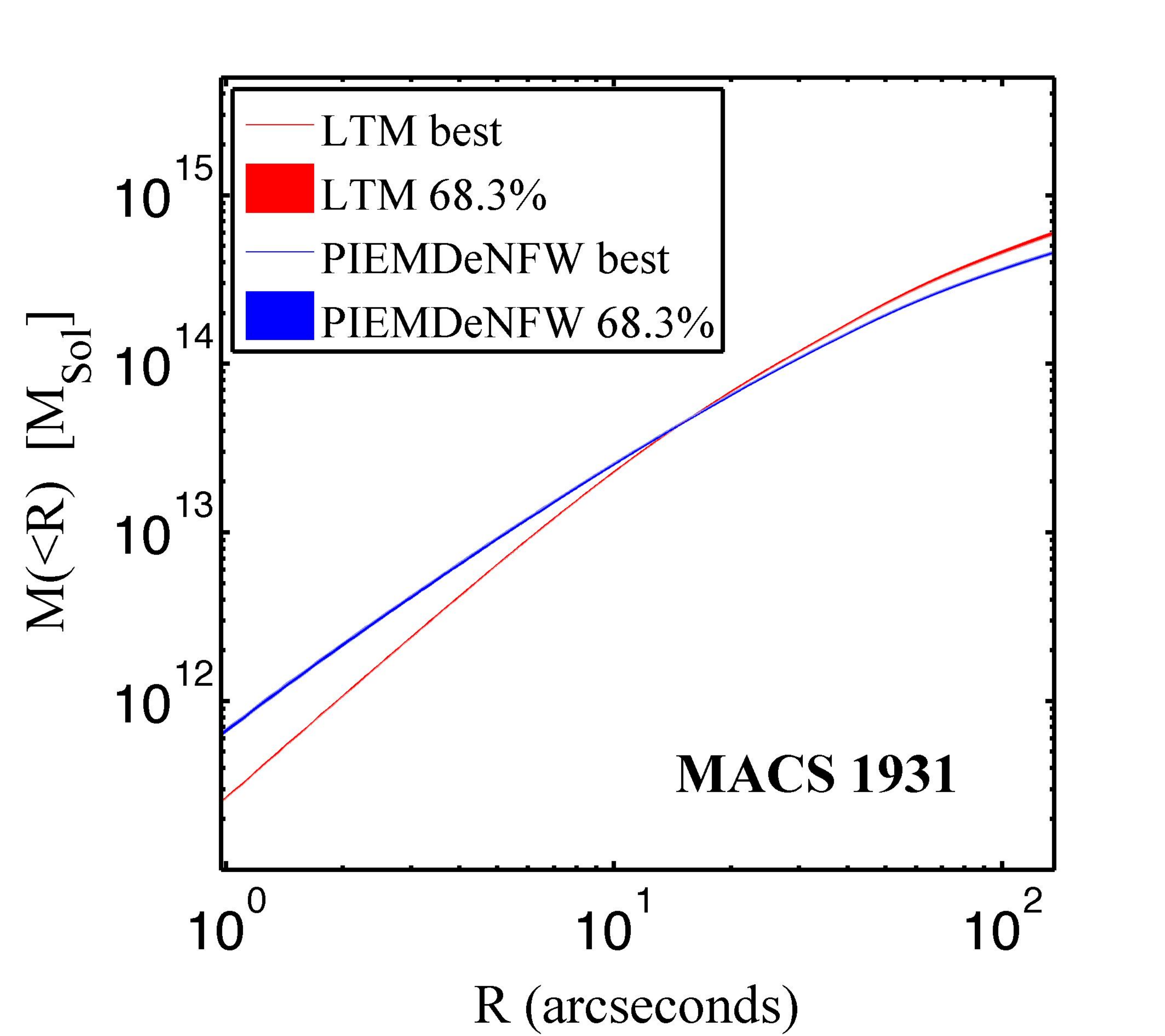}
\caption{Resulting 2D integrated mass profile as a function of radius, for an example cluster from our sample (MACS 1931; seen in Fig. \ref{curves0}), from both the LTM and PIEMDeNFW models (see \S \ref{Modeling}). Similarly, profiles for all other 24 CLASH clusters are shown in Fig. \ref{tryprof1} placed at the end of this manuscript.}\label{prof0}
\end{figure}

As above (\S \ref{LTM}), galaxies can be left freely weighted to be optimized by the model, and the minimization is similarly performed via a long MCMC with the same $\chi^2$ definition (eq. \ref{chi2SL}).

\subsection{Weak Lensing Regime}
To simultaneously fit for the strong and weak lensing regimes, we add a term for the total $\chi^2$ accounting also for WL data, so that the total $\chi^{2}$ is given by:
\begin{equation}\label{chi2tot}
\chi_{tot}^2=\chi_{SL}^2+\chi_{WL}^2
\end{equation}

with:

\begin{equation}\label{chi2WL}
\chi_{WL}^2=\sum_{i} \frac{(g^{'}_{1,i}-g_{1,i})^2+(g_{2,i}^{'}-g_{2,i})^2}{\sigma_{ell}^2} ,
\end{equation}

where $g_{1,i}$ and $g_{2,i}$ are the two components of the measured, complex reduced shear, of the $i$'th galaxy; $g^{'}_{1,i}$ and $g^{'}_{2,i}$ are the same two components as predicted by the model at each galaxy location; and $\sigma_{ell}$ is the width of the distribution of measured ellipticities (which governs the error in reduced shear), typically $\sim0.3$ which we adopt here as our nominal value following measurements of the standard deviation of a few input, shape measurement catalogs \citep[see also][]{Chang2013Wlsigma0p26,Newman2013}. While it is sometimes accustomed to use each background galaxies in the WL catalog individually, by using specifically the SNR and photo-$z$ of each galaxy, our tests while constructing the lens-modeling pipeline indicated that this has a negligible effect on the results, compared with using the fixed intrinsic ellipticity scatter we measured from our catalogs, or adopting the mean lensing depth, as we do here (see also \citealt{Newman2013} for a similar conclusion).

We disregard, i.e. we do not take into account in the $\chi^2$, galaxies for which the sign of the magnification by the lens model is negative, meaning that they lay inside the critical curves for the mean effective redshift of the WL sample. Also, one has to bear in mind that for background galaxies close to the cluster center, shape measurements may be affected by higher-order terms such as Flexion. It is currently uncertain by how much Flexion may actually affect one's shape measurements, a subject worth of proper investigation in future studies. We note, however, that in \citet{Merten2014CLASHcM} the effect of flexion on shape measurements in our HST WL catalogs was found to be negligible  - by comparing mass profiles constructed with and without the inclusion of background galaxies close to the cluster center, finding these are in excellent agreement. In addition, the RRG pipeline used here for shape measurements was found to correctly measure the reduced shear, to within 1\%, well into the SL regime \citep{MasseyGoldberg2008}.

For completeness, we will use throughout also the $reduced$ $\chi^{2}$, and note that the overall number of degrees-of-freedom (DOF) in the lensing model is:
\begin{equation} \label{DOF}
DOF= N_{SL,c}+N_{WL,c}-N_{p},
\end{equation}
where $N_{p}$ is the number of free parameters in the modeling, $N_{SL,c}$ is the number of SL constraints, and $N_{WL,c}$ the number of WL constraints. $N_{SL,c}$ is given by:
\begin{equation} \label{Nc}
N_{SL,c}= \nu(N_{im}-N_{s}) ,
\end{equation}

where the number of dimensions is $\nu=2$, since each image (and source) are characterized by two measures (e.g., x and y), $N_{im}$ is the total number of images used for the fit, and $N_{s}$ the number of systems (or sources; see also \citealt{Kneib1993Lenstool} for an equivalent formalism, or for an extension of the above to cases in which redshifts of background objects are also left to be freely optimized). From similar considerations, $N_{WL,c}$ is simply twice the number of galaxies used as constraints from our shape catalogs.

\tabletypesize{\tiny}
\begin{deluxetable*}{|c|c|c|c|c|c|c|c|c|c|c|c|c|c|c|}
\tablecaption{Summary of Analysis Results\label{ResultsTable}}
\tablehead{
Cluster$_{model}$ & Free Gals & Free Reds & $\chi^{2}_{SL}$ & $\chi^{2}_{WL}$ & $\chi^{2}/DOF$ & $N_{p}$ & $N^{c}_{SL}$ & $N^{c}_{WL}$ & $z_{WL}$ & rms & $\ln(ev)$ & $\theta_{e}$ & $M_{e}$ & $M_{2D}(< \theta\simeq2.3')$
}
\startdata
A209$_{LTM}$ &  1 & 2 & 117.2 & 1268.1 & $1385.3/1609=0.86$ & 11 & 8 & 1612 & 0.95& 2.05 & -694.6 & 9.0 & 0.82 & 1.97\\
A209$_{NFW}$ &  1 & 2 & 43.4 & 1262.9 & $1306.4/1611=0.81$ & 9 & 8 & 1612 & 0.95& 1.25 & -648.8 & 8.9 & 0.75 & 2.19\\
\hline
A383$_{LTM}$ &  2 & 0 & 152.8 & 1333.7 & $1486.5/1602=0.93$ & 10 & 22 & 1590 & 1.14& 1.50 & -705.0 & 15.2 & 2.13 & 2.39\\
A383$_{NFW}$ &  2 & 0 & 93.7 & 1330.0  & $1423.6/1600=0.89$ & 12 & 22 & 1590 & 1.14& 1.17 & -679.3 & 15.0 & 2.08 & 1.60\\
\hline
A611$_{LTM}$ &  1 & 1 & 88.8 & 836.0 & $924.8/1111=0.83$ & 9 & 26 & 1094 & 0.86& 1.14 & -460.4 & 18.9 & 4.70 & 5.67\\
A611$_{NFW}$ &  1 & 1 &  58.3 & 836.9 & $895.2/1112=0.81$ & 8 & 26 & 1094 & 0.86& 0.93 & -445.4 & 17.2 & 4.08 & 3.45\\
\hline
A1423$_{LTM}$ &  1 & 0 & 1.6 & 1286.2 & $1287.8/1603=0.80$ & 7 & 2 & 1608 & 0.92& 0.45 & -690.4 & 19.6 & 3.85 & 3.00\\
A1423$_{NFW}$ &  0 & 0 & 3.4 & 1284.0 & $1287.4/1604=0.80$ & 6 & 2 & 1608 & 0.92& 0.65 & -660.9 & 17.6 & 3.24 & 1.78\\
\hline
A2261$_{LTM}$ &  1 & 6 & 76.7 & 1119.9 & $1196.6/1440=0.83$ & 16 & 20 & 1436 & 0.79& 1.06 & -596.9 & 22.9 & 5.65 & 4.05\\
A2261$_{NFW}$ &  1 & 6 &  50.6 & 1121.8 & $1172.4/1441=0.81$ & 15 & 20 & 1436 & 0.79& 0.86 & -573.1 & 23.4 & 6.04 & 3.43\\
\hline
CL1226$_{LTM}$ &  2 & 1 & 131.4 & 2398.7 & $2530.1/1686=1.50$ & 10 & 20 & 1676 & 0.99& 1.53 & -1131.5 & 14.5 & 8.09 & 8.14\\
CL1226$_{NFW}$ &  2 & 2 &  40.1 & 2400.5 & $2440.6/1680=1.45$ & 14 & 18 & 1676 & 0.99& 0.88 & -1071.6 & 14.1 & 8.73 & 9.77\\
\hline
M0329$_{LTM}$ &  2 & 1 &  69.9 & 861.4 & $931.3/987=0.94$ & 9 & 16 & 980 & 1.18& 1.26 & -436.9 & 22.8 & 10.32 & 7.73\\
M0329$_{NFW}$ &  3 & 1 &  91.6 & 862.0 & $953.6/982=0.97$ & 14 & 16 & 980 & 1.18& 1.44 & -449.0 & 25.4 & 13.37 & 7.09\\
\hline
M0416$_{LTM}$ &  7 & 11 & 425.3 & 955.0 & $1380.3/1122=1.23$ & 24 & 44 & 1102 & 1.16& 1.72 & -962.3 & 25.0 & 10.35 & 10.52\\
M0416$_{NFW}$ &  7 & 11 & 348.9 & 961.1 & $1310.0/1117=1.17$ & 29 & 44 & 1102 & 1.16& 1.56 & -678.1 & 26.7 & 12.42 & 8.50\\
\hline
M0429$_{LTM}$ &  1 & 0 & 35.2 & 1069.7 & $1104.9/1304=0.85$ & 8 & 10 & 1302 & 1.08& 1.12 & -559.4 & 15.1 & 4.15 & 7.66\\
M0429$_{NFW}$ &  1 & 0 &  5.4 & 1064.8 & $1070.2/1305=0.82$ & 7 & 10 & 1302 & 1.08& 0.44 & -516.4 & 16.2 & 4.97 & 3.78\\
\hline
M0647$_{LTM}$ &  2 & 4 & 133.5 & 1336.8 & $1470.2/1537=0.96$ & 13 & 22 & 1528 & 1.14& 1.40 & -755.2 & 26.3 & 17.24 & 11.21\\
M0647$_{NFW}$ &  2 & 4 & 448.4 & 1337.1 & $1785.4/1538=1.16$ & 12 & 22 & 1528 & 1.14& 2.57 & -891.0 & 26.4 & 19.18 & 10.36\\
\hline
M0717$_{LTM}$ &  10 & 9 &  1858.7 & 962.0 & $2820.7/1045=2.70$ & 25 & 62 & 1008 & 1.04& 3.18 & -1390.0 & $\simeq$55 & 233.85 & 20.31\\
\hline
M0744$_{LTM}$ &  2 & 3 & 56.1 & 1359.7 & $1415.8/1488=0.95$ & 12 & 16 & 1484 & 1.32& 1.00 & -717.7 & 25.3 & 20.70 & 14.15\\
M0744$_{NFW}$ &  1 & 3 &  164.8 & 1360.9 & $1525.7/1486=1.03$ & 14 & 16 & 1484 & 1.32& 1.72 & -740.6 & 23.3 & 19.26 & 8.15\\
\hline
M1115$_{LTM}$ &  1 & 1 & 48.8 & 873.6 & $922.5/975=0.95$ & 9 & 12 & 972 & 1.03& 1.16 & -456.5 & 17.8 & 4.99 & 7.62\\
M1115$_{NFW}$ &  1 & 1 & 50.4 & 866.5 & $916.9/976=0.94$ & 8 & 12 & 972 & 1.03& 1.18 & -451.4 & 18.5 & 5.69 & 5.26\\
\hline
M1149$_{LTM}$ &  7 & 9 &  820.2 & 1458.7 & $2278.9/1706=1.34$ & 22 & 68 & 1660 & 0.99& 2.01 & -1072.5 & 20.4 & 9.83 & 14.36\\
\hline
M1206$_{LTM}$ &  5 & 5 & 335.5 & 989.4 & $1324.9/1203=1.10$ & 17 & 58 & 1162 & 1.13& 1.45 & -675.7 & 26.3 & 13.24 & 9.30\\
M1206$_{NFW}$ &  2 & 5 & 483.8 & 986.0 & $1469.8/1207=1.22$ & 13 & 58 & 1162 & 1.13& 1.74 & -695.4 & 27.3 & 15.31 & 9.26\\
\hline
M1311$_{LTM}$ &  1 & 0 &  9.6 & 758.6 & $768.2/892=0.86$ & 8 & 6 & 894 & 1.03& 0.69 & -369.7 & 13.5 & 4.24 & 8.96\\
M1311$_{NFW}$ &  1 & 1 & 2.8 & 760.3 & $763.1/892=0.86$ & 8 & 6 & 894 & 1.03& 0.37 & -361.6 & 14.8 & 5.09 & 5.66\\
\hline
M1423$_{LTM}$ &  1 & 1 & 111.4 & 1684.3 & $1795.7/1803=1.00$ & 9 & 28 & 1784 & 1.04& 1.21 & -824.1 & 17.6 & 7.55 & 10.93\\
M1423$_{NFW}$ &  1 & 1 & 164.8 & 1693.2 & $1858.0/1804=1.03$ & 8 & 28 & 1784 & 1.04& 1.47 & -853.6 & 17.8 & 8.20 & 6.13\\
\hline
RXJ1532$_{LTM}$ &  1 & 0 & 12.0 & 852.3 & $864.3/1004=0.86$ & 8 & 2 & 1010 & 1.07& 1.22 & -417.5 & 9.0 & 1.38 & 5.35\\
RXJ1532$_{NFW}$ &  1 & 0 &  1.4 & 850.7 & $852.1/1004=0.85$ & 8 & 2 & 1010 & 1.07& 0.41 & -411.5 & 10.5 & 1.85 & 3.09\\
\hline
M1720$_{LTM}$ &  1 & 4 & 83.9 & 1112.4 & $1196.3/1272=0.94$ & 12 & 20 & 1264 & 1.11& 1.15 & -642.4 & 20.4 & 7.22 & 7.15\\
M1720$_{NFW}$ &  1 & 4 & 253.9 & 1120.3 & $1374.2/1273=1.08$ & 11 & 20 & 1264 & 1.11& 1.99 & -643.6 & 19.8 & 7.36 & 3.35\\
\hline
M1931$_{LTM}$ &  1 & 2 & 249.2 & 797.2 & $1046.4/1422=0.74$ & 10 & 16 & 1416 & 0.82& 2.28 & -538.2 & 22.7 & 8.32 & 6.07\\
M1931$_{NFW}$ &  1 & 2 &  28.5 & 791.2 &  $819.6/1423=0.58$ & 9 & 16 & 1416 & 0.82& 0.77 & -461.0 & 21.8 & 7.82 & 4.57\\
\hline
M2129$_{LTM}$ &  1 & 6 & 560.0 & 1590.6 & $2150.7/1718=1.25$ & 16 & 32 & 1702 & 1.23& 2.42 & -950.4 & 19.2 & 9.23 & 11.85\\
M2129$_{NFW}$ &  1 & 6 & 333.1 & 1597.3 & $1930.4/1721=1.12$ & 13 & 32 & 1702 & 1.23& 1.86 & -846.7 & 21.8 & 12.99 & 8.59\\
\hline
MS2137$_{LTM}$ &  1 & 0 &  51.4 & 1619.6 & $1671.0/1572=1.06$ & 8 & 10 & 1570 & 1.12& 1.27 & -790.7 & 17.2 & 4.45 & 3.32\\
MS2137$_{NFW}$ &  1 & 0 &  19.1 & 1622.3 & $1641.3/1573=1.04$ & 7 & 10 & 1570 & 1.12& 0.77 & -779.3 & 17.0 & 4.42 & 2.69\\
\hline
RXJ1347$_{LTM}$ &  2 & 4 &  490.2 & 1246.3 &  $1736.6/1276=1.36$ & 14 & 24 & 1266 & 1.13& 2.61 & -738.1 & 33.3 & 22.65 & 18.04\\
RXJ1347$_{NFW}$ &  2 & 4 &  276.5 & 1251.3 & $1527.8/1274=1.20$ & 16 & 24 & 1266 & 1.13& 1.96 & -687.9 & 32.7 & 22.11 & 14.99\\
\hline
RXJ2129$_{LTM}$ &  1 & 4 &  88.5 & 1034.7 & $1123.1/1198=0.94$ & 12 & 18 & 1192 & 0.82& 1.26 & -522.8 & 13.3 & 1.85 & 3.39\\
RXJ2129$_{NFW}$ &  1 & 4 & 17.9 & 1036.4 & $1054.3/1199=0.88$ & 11 & 18 & 1192 & 0.82& 0.57 & -491.7 & 12.6 & 1.63 & 2.01\\
\hline
RXJ2248$_{LTM}$ &  1 & 14 & 318.8 & 923.2 & $1242.0/1216=1.02$ & 22 & 58 & 1180 & 1.12& 1.35 & -662.3 & 31.1 & 13.52 & 10.33\\
RXJ2248$_{NFW}$ &  1 & 14 &  547.0 & 942.7 & $1489.7/1217=1.22$ & 21 & 58 & 1180 & 1.12& 1.76 & -732.2 & 31.1 & 15.74 & 8.35
\enddata
\tablecomments{$\emph{Column 1:}$ abbreviated cluster name (see \S \ref{results} and \citet{PostmanCLASHoverview} for more cluster details), including each method used for the analysis (LTM or PIEMDeNFW, the latter being abbreviated here as ``NFW''; see \S \ref{Modeling} for details). $\emph{Column 2:}$ number of galaxies whose relative weight to the deflection map (i.e. its mass-to-light ratio) is left to be optimized by the minimization procedure. $\emph{Column 3:}$ number of background sources whose redshift was left to be optimized by the minimization procedure. $\emph{Column 4:}$ $\chi^{2}$ of the SL regime. $\emph{Column 5:}$ $\chi^{2}$ of the WL regime. $\emph{Column 6:}$ reduced $\chi^{2}$, $\chi^{2}/DOF$. $\emph{Column 7:}$ total number of free parameters in our models. $\emph{Column 8:}$  number of effective SL constraints. $\emph{Column 9:}$  number of effective WL constraints. $\emph{Column 10:}$ mean effective redshift of the weakly-lensed galaxies corresponding to the mean lensing depth $\langle\beta\rangle = \langle D_{\rm ls}/D_{\rm s} \rangle$ of the sample, defined as $\beta(\overline{z}_{\rm eff})=\langle\beta\rangle$. $\emph{Column 11:}$ image-plane reproduction $rms$ in arcseconds. $\emph{Column 12:}$ natural logarithm of the Bayesian evidence, calculated following the approximation given in \citealt{Marshall2006Evidence}. $\emph{Column 13:}$ Effective Einstein radius for $z_{s}=2$, in arcseconds ($\sqrt{A/\pi}$ where A is the area enclosed within the critical curves). $\emph{Column 14:}$ Mass enclosed within the critical curves for $z_{s}=2$, in $[10^{13} M_{\odot}]$. $\emph{Column 15:}$ 2D radially integrated mass within our analysis FOV, $\theta\simeq136\arcsec$, in $[10^{14} M_{\odot}]$ (see Figs. \ref{prof0} and \ref{tryprof1}).}
\end{deluxetable*}

\section{Individual Cluster Analysis}\label{results}

In this Section we briefly introduce each cluster in our sample, its lensing analysis, and notable results. Other technical or fitting results are summarized in Table \ref{ResultsTable}, and the resulting maps are explicitly shown in the figures throughout this work.

\subsection{Abell 209}

The galaxy cluster Abell 209 ($z=0.21$) is part of the main, relaxed cluster sample of the CLASH program. We found no record of a previous SL analysis of this cluster, nor any identification of multiple images outside the CLASH framework. Abell 209 has been, however, subject of various weak lensing studies (\citealt{Dahle2002, Smith2005, Paulin2007, Okabe2010}, see also \citealt{Merten2014CLASHcM,Umetsu2014CLASH_WL} for the analysis of most of the CLASH sample).

Here, we find the first, seven multiple images in this cluster, corresponding to three systems, which we consider as a secure set of constraints for the modeling. Our analysis therefore puts first constraints on the inner mass distribution of this cluster. We find a rather small lens, with an effective Einstein radius of $\sim9\arcsec$ for $z_{s}=2$, accounting for the small number of multiple images seen (also, two of the three systems found are locally lensed by a bright cluster member close, $\simeq18\arcsec$ to the BCG). In our modeling we use the photometric redshift of system 1 as fixed, and leave the redshifts of systems 2 \& 3, as well as the relative weight of the BCG, to be optimized by the MCMC. In the LTM case we also left the PA of the BCG and its core radius to be freely weighted. Our WL shape measurements include 806 galaxies that lay outside the critical curves and are used for the WL constraints. The results are summarized in Table \ref{ResultsTable} and seen in the Figures throughout.

\subsection{Abell 383}
Abell 383 ($z=0.189$) was the first cluster we analyzed in the CLASH framework \citep{Zitrin2011CLASH383}. SL features used here were known from previous works, spanning the redshift range $z=[1,6]$ \citep[e.g.][]{Sand2004,Newman2011,Richard2011A383highz}, supplemented with a few other multiple-images and candidates found by \citet{Zitrin2011CLASH383}. In particular, we have measured, in the framework of the CLASH-VLT campaign (PI: Rosati), a redshift for system 6 identified by \citet{Zitrin2011CLASH383}, to be at $z=1.83$. However, we did not use here candidate systems 7-9 uncovered by \citet{Zitrin2011CLASH383} since these were considered somewhat less secure; but we note that they are easily reproducible by the models and can therefore be considered secure hereafter for future analyses. In our LTM minimization we also leave to be freely weighted the two BCGs, and the BCG ellipticity parameters, and in our PIEMDeNFW modeling we also leave the secondary BCG ellipticity free. Our WL shape measurements include 795 galaxies that lay outside the critical curves and are used for the WL constraints, with an effective redshift of 1.14. In addition, we note that A383 is also one of the three CLASH clusters found to strongly magnify a background SN \citep{Patel2014SN,Nordin2014SN}, although this was not used here as a constraint.

\subsection{Abell 611}
The galaxy cluster Abell 611 ($z=0.29$) is a well studied, X-ray bright but relaxed cluster, with various previous lensing analyses \citep[e.g.][and references therein]{Richard2010locuss20,Newman2009,Donnarumma2011}. Three secure multiple image systems are known for this cluster, and we follow here the SL constraints (including revised redshifts) as given in \citet{Newman2013}. We also agree with their identification of additional central images for systems 1 and 3. As system 3 has no spectroscopic measurement, we leave its redshift to be optimized by the minimization procedure. We also leave the BCG to be freely weighted by the MCMC. 547 galaxies lay outside the critical curves and were used as the WL constraints, with an effective redshift of 0.86. 

\subsection{Abell 1423}
The galaxy cluster Abell 1423 ($z=0.213$) is part of CLASH's relaxed sample, and we found no record of a lensing analysis of this cluster outside the CLASH framework \citep[e.g.][]{Merten2014CLASHcM,Umetsu2014CLASH_WL}. We identify here a medium-to-small lens with a very rough Einstein radius of $\sim10-15\arcsec$. We do not find any secure multiply-imaged system but do uncover 2-3 candidate systems. We used one of them to construct a preliminary model using both parameterizations. This should be considered as a \emph{crude}, not well-constrained model due to the lack of multiple images. 804 galaxies lay outside the critical curves and were used as the WL constraints, with an effective redshift of 0.92. 

\subsection{Abell 2261}

The galaxy cluster Abell 2261 ($z=0.225$) has been subject to WL analyses based on Subaru data \citep[e.g.][]{Umetsu2009, Okabe2010}. \citet[][see also references therein]{Coe2012A2261}, constrained the inner mass profile of A2261 carrying out the first extensive SL analysis of this cluster using the 16-band HST imaging obtained as part of CLASH program, where multiple-images and candidates were uncovered with the aid of a preliminary LTM model (for another recent analysis see also \citealt{Ammons2014}). We use here the more secure identifications of the \cite{Coe2012A2261} list, as listed in Table \ref{multTable}, along with our WL catalog, to constrain the models. 

\subsection{CL J1226.9+3332}

The galaxy cluster CL J1226.9+3332 at $z=0.89$ is one of the hottest, most X-ray luminous systems at $z>0.6$ known to date \citep{Maughan2007}. In addition, \citet{JeeTyson2009} performed a weak lensing analysis of this cluster using HST/ACS images and found that this is also one of the most massive clusters known at $z>0.6$. However, we found no record of previous strong lensing analysis of CL 1226. In this work we find the first $\sim$15 multiple images and candidates corresponding to four background objects. One of these systems seems to be a very red giant arc prominent in the near-IR data. Our analysis also reveals a second central mass (sub-) clump, requiring a second DM eNFW halo with the PIEMDeNFW parametrization. 

\subsection{MACS J0329.6-0211}

In \citep{Zitrin2012CLASH0329}, we performed the first SL analysis known for the M0329 ($z=0.45$), finding six systems of multiple-images and candidates. One of the galaxies is a four-time imaged $z\sim6.2$ galaxy, whose properties were studied in \citet{Zitrin2012CLASH0329}. A spectroscopic redshift measurement for one of the systems uncovered was taken shortly thereafter, and we use it here as a constraint: system 2 is measured in the CLASH-VLT campaign to be at $z_{spec}=2.14$, very similar to the photometric redshift estimate used in \citet{Zitrin2012CLASH0329}, $z\sim2.17$.

\citet{Christensen2012Specs} performed spectroscopic observations for the $z\sim6.2$ galaxy, yet no secure determination of the spectroscopic redshift was achieved due to lack of emission lines. For the minimization procedure we only use systems 1-3, which we consider as most secure, where the redshift of system 3 is left free to be optimized by the models. For the LTM model, the two brightest galaxies are left to be optimized by the model, whereas for the PIEMDeNFW model, we use two eNFW halos centered on the two brightest galaxies, and leave the weight of three bright galaxies to be optimized by the model. 

\subsection{MACS J0416.1-2403}
M0416 ($z=0.40$) was first analyzed by \citet{Zitrin2013M0416} in the framework of the CLASH program, using both the LTM and PIEMDeNFW methods, and where we uncovered 70 multiple images of 23 background sources, and revealed an efficient, elongated bimodal lens. M0416 was then chosen as one of the HFF targets now being observed to a much greater depth than CLASH. Zitrin supplied mass models for this cluster available online through the HFF webpage, in the framework of the HFF map making campaign (PIs: Zitrin \& Merten) in which five different groups have submitted high-end mass models for use by the community. More recently, \citet{Jauzac2014M0416} and \citet{Diego2014M0416} have both published, independently, refurbished mass model for M0416 finding many additional multiple images in the supplemented HFF data, and \citet{JohnsonTraci2014FF} have previously published their lensing models for the HFF including M0416 (see also \citealt{Coe2014FF,Richard2014FF,Gruen2014WLclusters}; for other models). \citet{Grillo2014_0416} have recently also produced a very accurate lens model for M0416, in which several redshifts from our CLASH-VLT campaign were presented. Here, we use the same set of constraints from \citet{Zitrin2013M0416} with a slight revision reflected in Table \ref{multTable}. Spectroscopic redshift for the giant arc (sys 1) was available from \citep{Christensen2012Specs}.

\subsection{MACS J0429.6-0253}

Although M0429 ($z=0.399$) has been subject to various X-ray studies \citep{SmithAllen2007, ComerfordNatarajan2007CMrelation, Maughan2008clusterEv, Allen2008Xray, MannEbeling2012evolution}, we did not find any previous SL, nor WL, analyses for this cluster outside the CLASH framework. We uncover here the first two multiply-image families: one is a multiply-imaged arc with a distinctive knot in its middle, with a photometric redshift of $z_{s}\sim3.9$, and the second, a four times imaged blob, with a photometric redshift of $z_{s}\sim1.74$. Due to the very good agreement in redshift estimate among the uncontaminated multiple images of system 2, and the dropout feature of system 1, we adopt these photometric redshifts as fixed in our modeling (also, the relative distance ratio only slightly changes for the redshifts involved). 651 galaxies lay outside the critical curves and were used as the WL constraints, with an effective redshift of 1.08.

\begin{figure*}
\centering
\includegraphics[width=\textwidth]{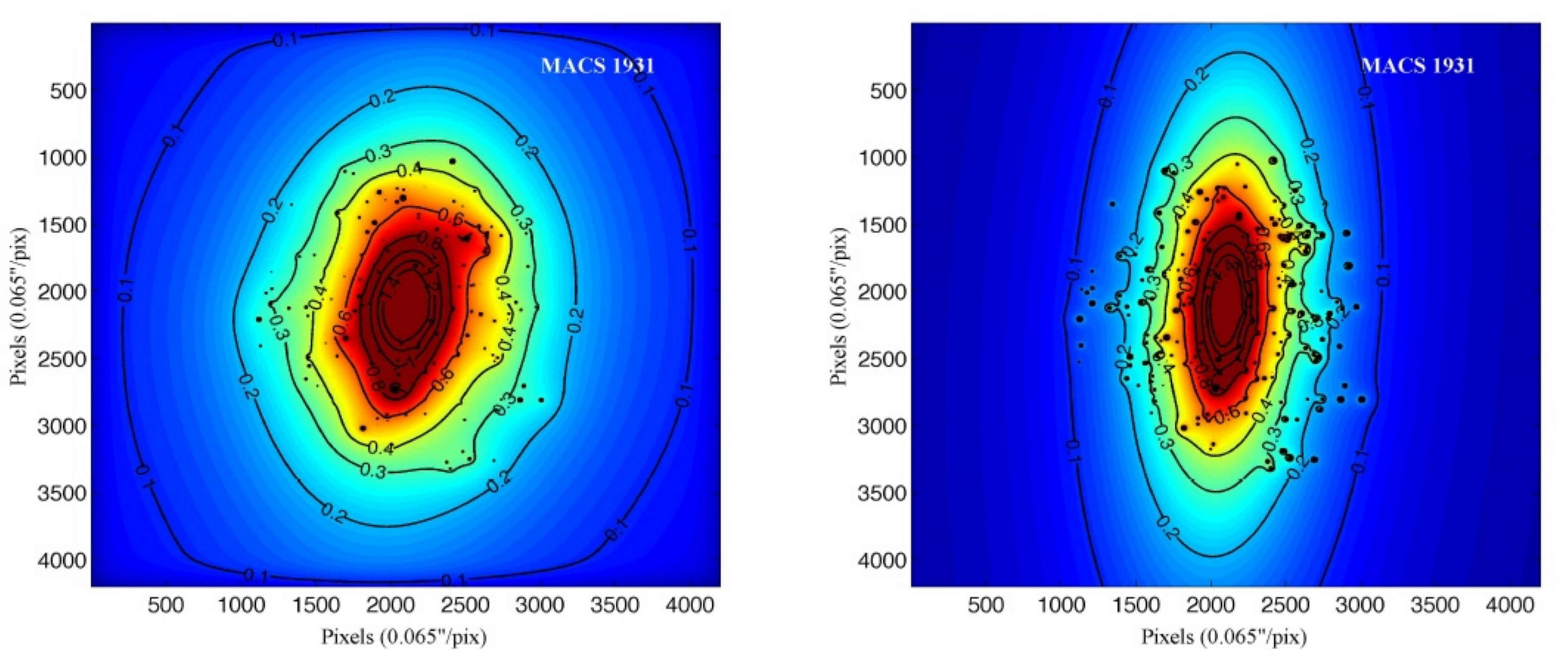}
\caption{Projected, surface mass density ($\kappa$) map from our LTM (\emph{left}) and PIEMDeNFW (\emph{right}) models for MACS 1931. Note the difference in ellipticity despite the similarity of the critical curves seen in Fig. \ref{curves0}. For similar maps of all other 24 CLASH clusters see Figs. \ref{Kaps1} and \ref{Kaps2}, respectively. These $\kappa$ maps are scaled to a fiducial redshift corresponding to $d_{ls}/d_{s}=1$ as was adopted for the CLASH and HFF mass model releases online.}\label{KapLTMNFW0}
\end{figure*}

\begin{figure}
\centering
\includegraphics[width=90mm]{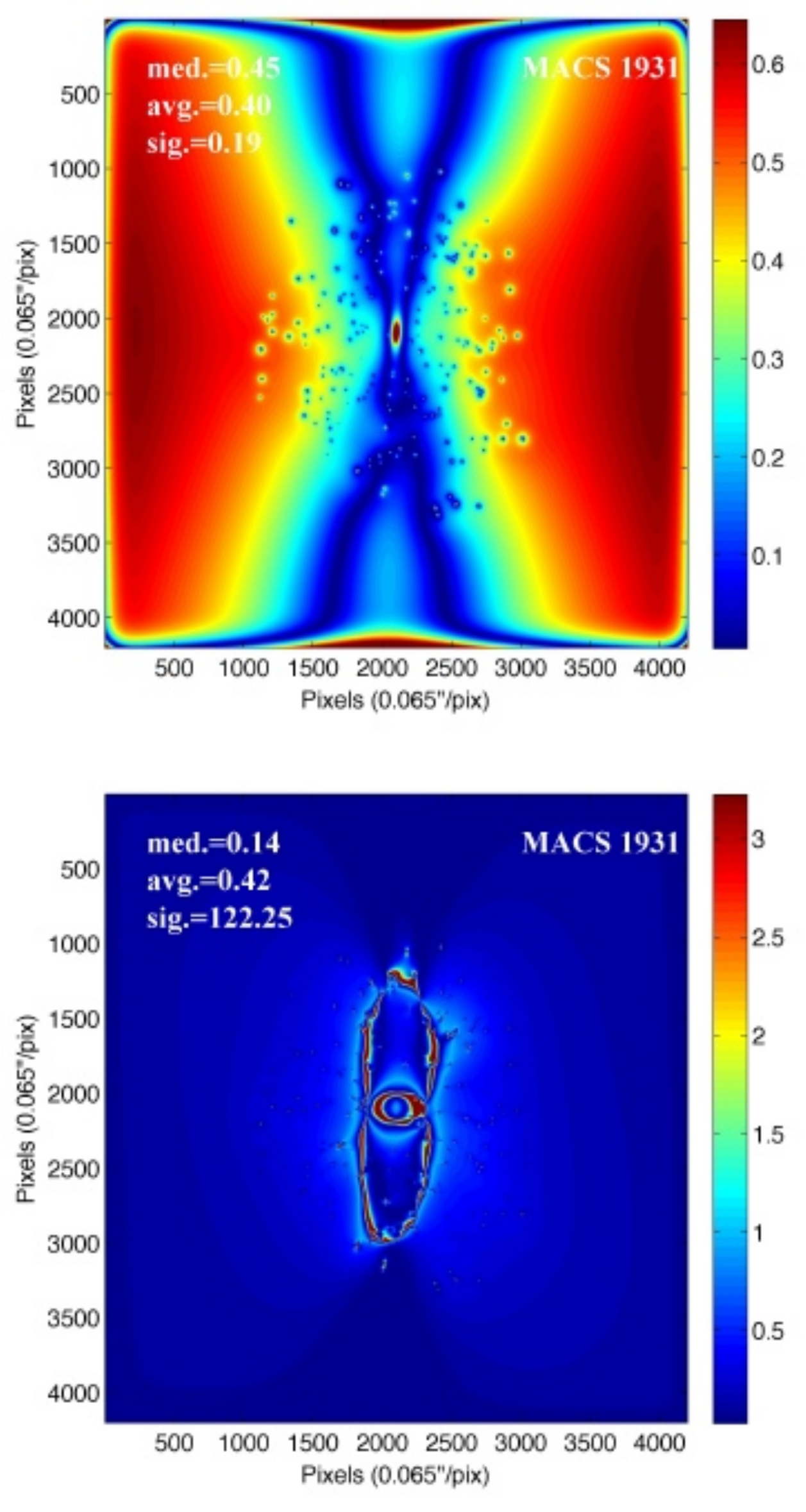}
\caption{\emph{Upper panel:} Absolute-value of the difference between the LTM $\kappa$ map and the PIEMDeNFW $\kappa$ map for MACS 1931, seen in Fig. \ref{KapLTMNFW0}, relative to the LTM map which we take as reference. On the Figure we note the average, median and standard deviation values. As can be seen, differences are mainly caused by ellipticities being assigned to the PIEMDeNFW mass density distributions directly, while the LTM mass density distribution simply follows the light and no ellipticity is introduced to it directly (albeit some ellipticity is incorporated in the BCG). Additionally, artifacts from the smoothing procedure introduce squareness in the LTM models near the edges of the FOV, contributing further to the discrepancy near the edges (this will be overcome in future analysis). A similar map for the other 24 CLASH clusters is shown in Fig. \ref{DifsKap1}. We find that the typical (\emph{median}) difference in $\kappa$, throughout this FOV among the CLASH sample, is $\sim40\%$, and the distribution of differences of all examined clusters is shown in Fig. \ref{hists}. See also \S \ref{discussion} for more details.\\
\emph{Lower panel:} Same as the \emph{Upper panel}, but now showing the absolute-value differences in magnification, relative to the LTM model. The majority of differences arise from the diverging critical curves and their surroundings, where farther away from them the error decreases. A similar map for all other CLASH clusters is shown in Fig. \ref{DifsMag1}. We find that the typical difference in $\mu$, throughout this FOV among our sample, is $\sim20\%$, and the distribution of differences is shown in Fig. \ref{hists}. See \S \ref{discussion} for more details.}\label{DifsKap0}
\end{figure}

\subsection{MACS J0647.7+7015}
M0647 ($z=0.591$) is part of the 12 MACS $z>0.5$ cluster sample \citep{EbelingMacs12_2007}, and as such was first analyzed by \citet{Zitrin2011_12macsclusters} in their work on this sample. \citet{Zitrin2011_12macsclusters} identified the first multiple images in this cluster, later revised and supplemented with additional images from CLASH data, revealing also a likely $z\sim11$ multiply-imaged galaxy \citet{Coe2012highz}, which is the highest-redshift galaxy candidate known to date. We used as constraints the secure identification listed in \citet{Coe2012highz}, as seen in Table \ref{multTable} here. 

\subsection{MACS J0717.5+3745}
M0717 ($0.546$) is also one of the 12 MACS clusters at $z>0.5$ \citep{EbelingMacs12_2007}, and as such was first analyzed by \citet{Zitrin2009_macs0717,Zitrin2011_12macsclusters} in their work on this sample. \citet{Zitrin2009_macs0717} found, using their LTM method, many multiple images in this cluster, which revealed a complex lens that constitutes the largest strong lens known to date, with an Einstein radius of $z\sim55\arcsec$. The high lensing power of this cluster, which is a notable part of its surrounding cosmic web \citep{Ebeling2004filament}, qualified it as well for the HFF program, with observations planned for the near future. Other mass models for this cluster were published, both in the SL and WL regimes \citep[e.g.][]{Jauzac2012WL0717,Limousin2012_M0717,Diego2014M0717,Richard2014FF,JohnsonTraci2014FF,Medezinski2013M0717}. In the latter work, we have also revised our initial multiple-image list from \citep{Zitrin2009_macs0717}, following \citet{Limousin2012_M0717} with additional corrections, as listed in Table \ref{multTable} here. In addition, \citet{Vanzella2014z6p4M0717} identified two spectroscopically confirmed $z=6.4$ lensed by M0717, which could be multiple images of the same background galaxy  - as was considered in some of the works mentioned above - yet we did not use these here as constraints. Note also that in the HFF framework, we have submitted two models for this cluster using both LTM Spline interpolation smoothing and a Gaussian smoothing as the one we use here. \citet{Limousin2012_M0717} have shown that when modeled with analytic DM halos, this cluster cannot be well modeled with one halo and needs five of them. For that reason, we do not use here the PIEMDeNFW parametrization and for the current work we remodeled the cluster only in the LTM Gaussian smoothing method, with the same pipeline as all 25 clusters (which was slightly refurbished since we made our HFF models). In this method the same simple procedure is applied to all scales, from galaxy-group lenses up to very complex clusters such as this one. Other studies of systematics in this cluster can be performed elsewhere, such as in the framework of the HFF.

\subsection{MACS J0744.8+3927}
As the two previous clusters in our CLASH list, M0744 ($z=0.698$) is also part of the 12 MACS clusters at $z>0.5$ \citep{EbelingMacs12_2007}, and as such was first analyzed in \citet{Zitrin2011_12macsclusters}, where the first multiply imaged galaxies known for this cluster were found. We have now revised our identification using CLASH imaging, and revealed several additional multiply-imaged galaxies that we use as constraints; see Table \ref{multTable} or Figs. \ref{curves1}-\ref{curves4} for more details. For our PIEMDeNFW, we use here two eNFW DM halos, as the constraints are not well explained by a single central halo. 742 galaxies lay outside the critical curves and were used as WL constraints, with an effective redshift of 1.32.

\subsection{MACS J1115.9+0129}
We found no record of previous lensing analyses of M1115 ($z=0.352$) outside the CLASH framework, and present here the first strong and weak lensing analyses in HST data, including the multiple images and candidates identification. In the SL regime our model includes two multiply-lensed systems. The first system includes a low surface brightness giant arc and its counter images, which was also noted and targeted spectroscopically by \citet{Christensen2012Specs}, yet no emission lines were found and thus no unambiguous redshift could be determined. \citet{Christensen2012Specs} have concluded a plausible redshift of either $z\sim0.5$ or $z\sim3.5$. For our lens models we adopt a fixed photometric redshift of 2.84 as obtained from the BPZ program in the CLASH pipeline. The second system is a small blob imaged three times, where its images seem to follow tightly the symmetry of the lens, and show similar colors and photometric redshifts. To allow for some flexibility we allowed the redshift of this system to be optimized by the minimization procedure. Our models suggest that the $z=2.84$ redshift we adopted to system 1 may be significantly higher than its true redshift (both models suggest system 2 lies at a significantly higher lensing distance than system 1). Correspondingly, the presented models have to be treated with somewhat more caution, warranting a future revision. 486 galaxies lay outside the critical curves and were used as the WL constraints, with an effective redshift of 1.03.

\subsection{MACS J1149.5+2223}
M1149 ($z=0.544$) is also one of the 12 MACS cluster sample at $z>0.5$ \citep{EbelingMacs12_2007}, and as such was first analyzed by \citet{Zitrin2009_macs11495,Zitrin2011_12macsclusters}, where several multiply-imaged galaxies were uncovered in this cluster including a giant multiply-imaged spiral \citep[see also][]{Smith2009M1149}. With CLASH data we have now revised our multiple image identification and revealed other multiple images \citep[e.g.][]{Zheng2012NaturZ}, which we use here as constraints; see Table \ref{multTable} or Figs. \ref{curves1}-\ref{curves4} for more details. \citet{Smith2009M1149} found that a single DM halo does not describe well the cluster, and three more DM halos had to be added by them to obtain an accurate fit. For that reason, we do not attempt to model this cluster with the PIEMDeNFW method here, and only concentrate on a new model using our LTM technique. In addition, due to its lensing capabilities, M1149 is also part of the HFF program. In that framework, we have submitted two models for this cluster using both a Spline interpolation smoothing and a Gaussian smoothing as the one we use here. For the current work we remodeled the cluster with the exact same pipeline as for all 25 CLASH clusters (which was slightly refurbished since we made our HFF models), only in the LTM Gaussian smoothing method. Other studies of systematics in this cluster can be performed elsewhere. For additional, recent models for this cluster see \citet[][]{Richard2014FF,JohnsonTraci2014FF,Rau2014M1149}. In our modeling we also use, as WL constraints, 830 galaxies that lay outside the critical curves, with an effective redshift of 1.32.

\subsection{MACS J1206.2-0847}
M1206 ($z=0.44$) was first analyzed by \citet{Ebeling2009} based on a prominent giant arc seen west of the BCG. Using CLASH imaging, and our LTM technique, \citep{Zitrin2012CLASH1206} have revealed 47 new multiple images and candidates of 12 background sources, for some of which spectroscopic redshifts were obtained in our CLASH-VLT campaign, allowing to constrain the mass profile. Our profile was found to agree well also with an independent WL analysis by \citealt{Umetsu2012}, and a dynamical analysis by \citealt{Biviano2013CLASH1206}. We use here a similar set of constraints to model the cluster in both parameterizations, now including, also, HST WL data. 581 galaxies lay outside the critical curves and were used as the WL constraints, with an effective redshift of 1.13.

\subsection{MACS J1311.0-0311}
We did not find a report of any strong or weak lensing analysis for M1311 ($z=0.494$) outside the CLASH framework. We identify here two multiply-image systems and an additional candidate system, and present the first analysis of this cluster. The first system is a dropout with photometric redshift of $z=5.82$, and the second has a photometric redshift of $2.40$. We keep these redshifts fixed throughout our minimization, while allowing the BCG's weight to be optimized, as for most of the clusters we analyzed. 447 galaxies lay outside the critical curves and were used as the WL constraints, with an effective redshift of 1.03.

\subsection{MACS J1423.8+2404}
The galaxy cluster M1423 ($z=0.545$) was first analyzed by \citet{Limousin2010M1423} by SL+WL data together. \citet{Limousin2010M1423} found and spectroscopically measured two multiply-imaged systems that later \citet{Zitrin2011_12macsclusters} used in their analysis of the 12 $z>0.5$ MACS clusters \citep[see][]{EbelingMacs12_2007}. Using CLASH data, we have now found an additional multiply imaged system comprising five multiple images (system 3 here), which we use here as additional constraints. Additionally, we have found \citep[see also][]{Bradley2013highz} a few higher-redshift candidates which are possibly multiply imaged at $z\sim6-7$ (candidate system 4 here). This option should be investigated more thoroughly in future studies.

\subsection{RXJ1532.9+3021}
RXJ1532, which we also refer to as MACS 1532 \citep{Ebeling2010FinalMACS} contains a remarkable star-forming BCG residing in a cool-core cluster. We found no record of a previous lensing analysis prior to CLASH data. M1532 is also one of the three CLASH clusters found to strongly magnify a background SN \citep{Patel2014SN,Nordin2014SN}. The latter works have also presented models for this cluster for determining the background SN magnification, however no multiple images were listed therein. We do not find any secure set of multiple images in this cluster, and have only identified one candidate system. We correspondingly present \emph{crude and preliminary} mass models for this cluster, while including also the WL information as constraints. 505 galaxies lay outside the critical curves and were used as the WL constraints, with an effective redshift of 1.07.

\subsection{MACS J1720.2+3536}
Also M1720 ($z=0.387$) is one of the three CLASH clusters found to strongly magnify a background SN \citep{Patel2014SN,Nordin2014SN}. \citet{Nordin2014SN} presented a mass model (or magnification map) for this cluster, yet did not list the multiple images used as constraints. \citet{Patel2014SN} also presented a model, based on WL data and on the list of multiple images we have now identified and list here for the first time (Table \ref{multTable}).

\subsection{MACS J1931.8-2635}
We did not find any record of a previous lensing analysis for M1931 ($z=0.352$) outside the CLASH framework. We identified 22 new multiple images and candidates of 7 background sources, which we use as SL constraints on top of our WL shape measurements, revealing a remarkably elongated lens. 708 galaxies lay outside the critical curves and were used as the WL constraints, with an effective redshift of 0.82. 

\subsection{MACS J2129.4-0741}
M2129 ($z=0.59$), like several other clusters mentioned above, is also one of the 12 MACS clusters at $z>0.5$ \citep{EbelingMacs12_2007}, and as such was first analyzed in \citet{Zitrin2011_12macsclusters}, where the first multiple images in this cluster were uncovered. Here, we supplement this identification with additional multiple images now uncovered in CLASH data, and use these to constrain the cluster lens model, in conjunction with the WL data. 851 galaxies lay outside the critical curves and were used as the WL constraints, with an effective redshift of 1.23.

\subsection{MS 2137-2353}
MS2137 ($z=0.313$) seems to be a well relaxed cluster and exhibits a giant arc. Several attempts to model the mass in cluster took place albeit with some tension between the results \citep[e.g.][and references therein]{Gavazzi2003,Sand2008,Donnarumma2009MS2137,Newman2013}. The tension mainly arises from different mass profile estimates and is due to lack of enough multiple images to properly constrain the slope of this cluster: only two multiply-imaged galaxies were known before, the aforementioned giant arc and an additional system, both at a similar redshift of $z_{s}=1.5$ \citep[][and references therein]{Donnarumma2009MS2137}. Since the mass profile is coupled to the lensing distance ratio between different-redshift multiply lensed galaxies, it was essentially impossible to place strong constraints on the inner mass profile of this cluster from lensing alone. Using our LTM method we were now able to identify three images of an additional multiply-lensed galaxy, verified by the CLASH HST imaging and VLT spectroscopy from the CLASH-VLT run (PI: P. Rosati), which allows to reliably constrain the inner mass profile of this cluster (since the latter system has a different, and substantially higher redshift than the first two systems, $z_{s}=3.09$). 785 galaxies lay outside the critical curves and were used as the WL constraints, with an effective redshift of 1.12.

\subsection{RXJ1347.5-1145}
RXJ1347 ($z=0.45$) is one of the most X-ray luminous clusters known \citep[e.g.][as one example]{Schindler1995RXJ1347}, and as such was the subject of several lensing analyses \citep{Halkola2008RXJ1347,Bradac2008rxj1347,KohlingerSchmidt1347}. Using our LTM method we chose the most reliable identification of multiple images from these previous lensing works, listed in Table \ref{multTable}, as constraints for our model along with the HST WL data. 633 galaxies lay outside the critical curves and were used as the WL constraints, with an effective redshift of 1.13.

\subsection{RXJ2129.7+0005}
RXJ2129 ($z=0.234$)  was previously studied in the framework of the LoCuSS collaboration: A WL analysis was published for example by \citet{Okabe2010}, and a SL analysis was published by \citet{Richard2010locuss20}, based on one identified system. The redshift of this system was published by \citet{Richard2010locuss20} to be $z_{s}=1.965$. Recently, \citet{Belli2013Clusters} revised the (spectroscopic) redshift measurement to $z=1.522$, which is the redshift we adopted for our analysis. In addition, we publish here five new sets of multiple images and candidates, whose redshifts we left to be optimized by the minimization procedure. These allow us to put much stronger constraints, for the first time, on this cluster's mass distribution and profile. 

\subsection{RXC J2248.7-4431}
RXJ2248 ($z=0.348$), also known as Abell 1063S, was recently analyzed for the first time by CLASH \citep[][see also \citealt{Balestra2013_z6_2248,Gruen2013RXJ2248WL}]{Monna2014RXC2248}, uncovering many multiple images including a $z\sim6$ galaxy imaged five times. We use these constraints as listed in Table \ref{multTable} for the SL part, jointly with the HST WL shape measurements. Note also that we have already published mass models for this cluster (SL only) in the framework of our HFF map-making group (PIs: Zitrin \& Merten) available online, yet here we rerun those models with the slight modifications to our code to be coherently analyzed as all other clusters in our sample. Also, other mass models from other HFF lens modelers are available online through the HFF page (e.g. \citealt{JohnsonTraci2014FF,Richard2014FF}).

\section{Results and Discussion}\label{discussion}

We constructed lens models for the full CLASH cluster sample, using both the SL, and WL signals, in deep HST observations taken in the CLASH program. For most clusters, we used \emph{two distinct} common parameterizations: the full LTM parametrization assuming LTM for both galaxies \emph{and} the DM, and a parametrization in which LTM is only assumed for the galaxies, while the DM is modeled separately and analytically with an eNFW halo (or two, for more complex merging clusters). Our main goal was to present here the mass models, and the multiple image catalogs, and to release them to the community along with an investigation of the typical, systematic differences.

In Figs. \ref{curves0}, and \ref{curves1}-\ref{curves4} we plot the critical curves for all clusters, from the two parameterizations, on an RGB image constructed for each cluster using the 16-band CLASH observations. In Figs. \ref{prof0} and \ref{tryprof1} we plot the two (where available) resulting 2D-integrated profiles for all clusters. Figs. \ref{KapLTMNFW0} and \ref{Kaps1}-\ref{Kaps2} show the dimensionless mass density distributions $\kappa$ for the LTM models and PIEMD+eNFW models, respectively. The differences in $\kappa$ and magnification between the two models, relative to the LTM model (which we arbitrarily chose as our reference model), are seen in Figs. \ref{DifsKap0} and \ref{DifsKap1}-\ref{DifsMag1}, respectively. Note that the ordering in Figs. \ref{curves1}-\ref{DifsMag1} is similar, yet slightly differs from the order in \S \ref{results}.

\subsection{Systematics and Statistical Uncertainties}\label{errorsSec}
We discuss the statistical and systematic uncertainties probed by our analysis with two distinct mass modeling parameterizations.

\subsubsection{Magnification and Mass-Density Maps}\label{mukappaSysSec}
As is evident from Figs. \ref{curves0}, and \ref{curves1}-\ref{curves4}, the critical curves from both parameterizations (where available), for each cluster, are in overall good agreement. Despite the quite distinct parameterizations, this may not be too surprising as in practice similar multiple-image constraints are used for both solutions, directly determining where the critical curves should pass for each multiply imaged system. This is also why it is important to compare the resulting maps to one another, in order to see what are the differences both in the SL regime and across the larger FOV where the constraints from SL are poor on non-existent.

We wish to provide a reasonable estimate of how strongly the choice of parametrization affects the resulting maps of mass density, $\kappa$, and magnification, $\mu$. Such an estimate of the underlying systematics is crucial for any work which relies on the lens models for their study, such as measurements of the actual lensed volumes, properties of lensed galaxies, or the intrinsic properties of magnified, high-redshift background objects, especially in the current epoch of increasing interest in magnified, high-redshift galaxies and with recent, extensive cluster lensing surveys using HST such as LoCuSS (PI: G. Smith; e.g. \citealt{Richard2010locuss20}), CLASH \citep{PostmanCLASHoverview}, and the ongoing HFF \citep{Lotz2014AAS_FFreview}. For each cluster, for the $\kappa$ and $\mu$ maps separately, we subtract each map from the corresponding map in the other parametrization, and divide the absolute-value of the result by the LTM map, as reference, to obtain relative residual maps. These are seen for one example cluster in Fig. \ref{DifsKap0}, and for the other 24 clusters - as stamp images in Figs.  \ref{DifsKap1} and \ref{DifsMag1}, respectively. For each cluster we also note therein the median, mean, and $1\sigma$ dispersion of each map.

Regarding the magnification differences, a few things are evident from Figs. \ref{DifsKap0} and \ref{DifsMag1}. First, it is clear that most of the relative systematic differences are seen next to the critical curves, where the magnification diverges. Second, as a result, the median and mean differ significantly in most cases, and the standard deviation is huge. This is not surprising, as these are governed by the error induced by the diverging critical curves. In Fig. \ref{hists}, we make a histogram of both the absolute and the relative difference in the magnification, gathered from all pixels across the 23 clusters that were analyzed with the two methods. The 68.3\% C.L. of the relative magnification differences is [0.08,0.44], with a median(mean) of 0.22(0.27), implying a typical $\sim20\%$ systematic error relative to the reference LTM model, on the magnification across the probed, central $\sim[4.6'\times4.6']$ FOV. In terms of absolute magnification differences, the 68.3\% C.L. in $\Delta\mu$ is [0.11,1.12], with a median(mean) of 0.37(0.65).

Interesting information is also gained by looking at the $\emph{kappa}$ map relative differences. As seen from Figs. \ref{DifsKap0} and \ref{DifsKap1}, the typical $\emph{kappa}$ relative differences are much larger than those in the magnification. This may be surprising at a first glance, as we know that the magnification varies more rapidly than $\emph{kappa}$, and is more susceptible to small changes in $\emph{kappa}$ (recall that $\mu=1/((1-\kappa)^{2}-\gamma^{2})$). However, since each multiple-image system supplies direct constraints on the position of the critical curves, yet only constrains the total mass within those critical curves, the constraints on the distribution of the magnification are, in a way, more direct than those on kappa. More importantly, while in the PIEMD+eNFW parametrization the ellipticity is embedded directly into the mass distribution, in the LTM case there is no ellipticity assigned to the mass distribution (although some ellipticity is embedded in the BCG); the overall ellipticity, other than that induced by the BCG, only enters in the form of an external shear that has an effect on the magnification map or critical curve's ellipticity (and on the shear map), yet it does not affect the mass distribution itself, which is coupled to the light. This often creates a prominent discrepancy between the mass density distributions resulting from these two parameterizations: since lensing only constrains combinations of $\kappa$ and $\gamma$, it is possible to reproduce similar critical curves (or magnification maps and to some extent, reduced shear maps) from these two different parameterizations that have distinct $\kappa$ maps, and the degeneracy between them is not broken with typical SL+WL lensing constraints alone. In other words, even the combination of SL and WL does not seem to be enough to distinguish between a model in which all the lensing signal is attributed to an intrinsically elliptical lensing cluster, and a model in which no overall ellipticity is input into the mass distribution and is only imitated by adding an external shear \citep[see also][]{MatthiasBartelmann1995Ellip}. This degeneracy might be broken, in principle, using these lensing constraints together with additional, independent and direct constraints on the magnifications (e.g such as lensed supernovae Ia, albeit these are rare), or relative magnifications between multiple images of the same source. One possibility to further examine this, would be to construct a model that allows both for intrinsic ellipticity \emph{and} external shear, to characterize better the degeneracy between them and see what are their relative contributions, on an ensemble of clusters. Additionally, a comparison to numerical simulations may also shed some light on the true underlying mass distribution of such clusters. It will be worth pursuing such paths in the near future.

Another minor contribution to the systematic differences in $\emph{kappa}$ comes from a numerical artifact in the LTM method that induces squareness into the $\emph{kappa}$ map close to the edges of the FOV, due to imperfect (or aperiodicity in the) boundary conditions for the Fourier transform used in our smoothing procedure (significantly speeding up the calculation). This is seen clearly in the patterns shown in Figs. \ref{Kaps1} and \ref{DifsKap1}, and can be overcome in the future by refining the boundary conditions, or, by simply taking a larger mock FOV to then be cut to the desired size while getting rid of the affected corners of the larger FOV (such a solution would be too time consuming, however, to be performed on the 25 cluster sample in a reasonable time frame for this work). In fact, we note that very recently we have managed to overcome this artifact and now produce ``cleaner" maps without affecting much the speed of the calculation. This, however, will only be implemented and better tested in future analyses. This artifact contribution here, though, seems to be very minor in most of the field compared to the intrinsic differences between the two parameterizations, and contributes significantly only very close to the edges (see e.g. \S \ref{massprofiles}).

\begin{figure*}
\centering
\includegraphics[width=160mm]{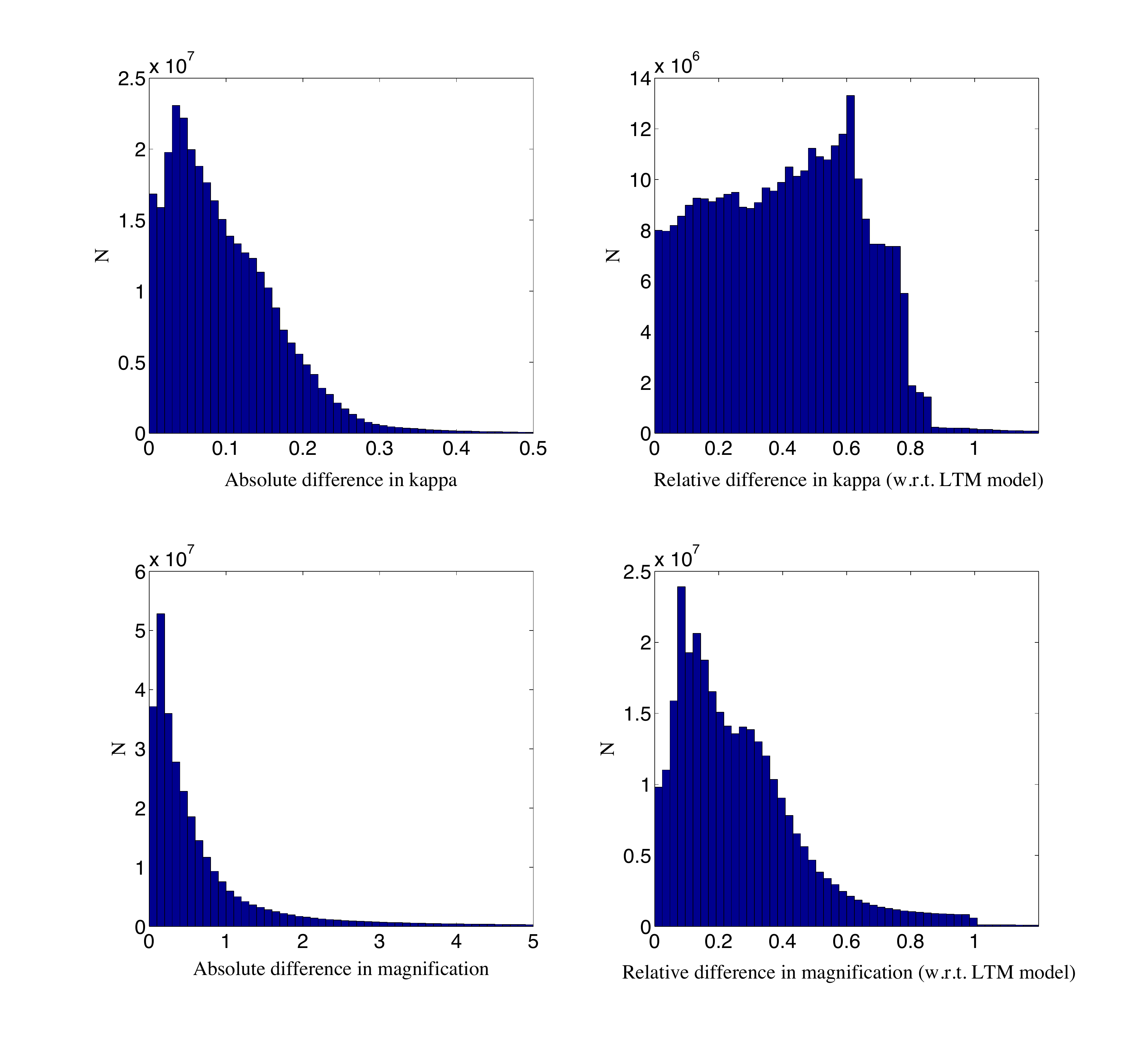}
\caption{Histograms of the absolute \emph{(left)}, and relative \emph{(right)}, differences in the surface mass density $\kappa$ (\emph{upper panel}), and the magnification $\mu$ (\emph{lower panel}), between the two methods we employed here, reflecting the systematics differences between them. For details see \S \ref{mukappaSysSec}.}\label{hists}
\end{figure*}

We show histograms of both the absolute and relative differences in $\kappa$, in Fig. \ref{hists}. The 68.3\% C.L. of the relative $\kappa$ differences is [0.14,0.65], with a median(mean) of 0.42(0.41), implying a typical $\sim40\%$, relative systematic error. In terms of absolute kappa differences, the 68.3\% C.L. in $\Delta\kappa$ is [0.03,0.17], with a median(mean) of 0.08(0.10). This typical, relative difference we find ($\sim40\%$), constitutes a significant systematic error, arising mainly from degeneracies inherent to lensing as aforementioned (embedded in the two different parameterizations). Although most lensing-related studies are more dependent on the magnification estimate or the overall mass-distribution properties than on the value of each point in the $\emph{kappa}$ map, so these errors may have less affect on related science, they are important to be aware of. 

Since the lens models are constrained using multiply-imaged sources at different redshifts, this places immediate constraints on the resulting mass profile, so that one expects smaller differences in the averaged $\emph{kappa}$ or enclosed mass $\emph{profile}$ (which is of course also relevant for related studies, such as structure formation, the concentration-mass relation etc.), especially within the SL regime or critical curves. We probed the resulting enclosed masses and integrated mass profiles in \S \ref{remeSec} and \ref{massprofiles} below.

Lastly, we note that the values mentioned above, extracted for the histograms of relative differences in the magnification and surface density, remain effectively unchanged whether we probe the $[4.6\arcmin \times4.6\arcmin]$ FOV, or an inner $[4\arcmin \times4\arcmin]$ FOV, showing the the effect of the boundary artifact discussed above on the overall differences, is very minor.

The examination of systematics in lens modeling has been a long-standing crucial task, especially in recent years where advanced modeling techniques have been developed and HST space imaging has allowed increasingly accurate lensing analyses and studies of high-redshift magnified galaxies. Our results, to our knowledge, constitute the first time such a systematic uncertainty estimate is performed over a meaningful sample of well-analyzed clusters, with two different common techniques. Other comparisons were made for single clusters in past studies \citep[e.g.][]{Zitrin2010_A1703,Coe2012A2261}. Our insight on the systematics can be regarded as an introduction to systematics in the HFF program in which 6 clusters were modeled using various different parameterizations or techniques, by five different lens-modeling groups, to assess the underlying systematic differences \cite[see for example][]{Coe2014FF}, which will also be tested on simulated clusters. In that case, more modeling methods will be used to assess the underlying systematics compared to our work here, but on a much smaller sample of clusters. Both tests are of course important to perform.

\subsubsection{Einstein Radius and Einstein Mass Distributions}\label{remeSec}
The effective Einstein radii and enclosed masses for a fiducial redshift of $z_{s}=2$ are summarized in Table \ref{ResultsTable}. Note that here we measured the effective Einstein radii numerically, by summing all pixels enclosed within the tangential critical curves according to the magnification sign and including also the area within the radial critical curves, where the sign of magnification flips again, to derive the critical area, A, where $\theta_{e}=\sqrt{A/\pi}$. The measured Einstein radii of the sample clusters, from the two distinct analyses, agree within $\sim10\%$, where the enclosed masses agree typically within $\sim15\%$ (but with some outliers). This is a good agreement that is not surprising: each multiply-imaged system directly places strong constraints on the Einstein radius for its redshift, and thus the enclosed mass (the mass enclosed within the Einstein radius, e.g. for a circularly symmetric lens, is proportional to the Einstein radius squared, $M_{e}\propto \theta_{e}^{2}$). We adopt therefore these values, i.e. $\sim10\%$ and $\sim15\%$, as the representative systematic uncertainties on the Einstein radius and enclosed mass, respectively.

\subsubsection{Mass Profiles}\label{massprofiles}
In Figs. \ref{prof0} and \ref{tryprof1} we show the resulting 2D integrated mass profiles of the CLASH sample, from our two modeling methods. The main difference between the two profiles is in each method's prescription: the PIEMDeNFW fit, governed by the analytic DM form, is bound to be well behaved and show a profile following, roughly, the input analytic (NFW in our case) form; while the LTM fit is not coupled to any analytic form, and although the mass $\emph{distribution}$ is coupled to the light distribution, the $\emph{profile}$ is in practice more flexible than the first method, in the sense that it does not follow a certain predetermined form, thus probing a wider range of profile shapes. We test the discrepancy in the total 2D-integrated mass between the two methods, $M_{2D}(<\theta\simeq136\arcsec$). We find a typical (median) $38\%$ difference between the two values over all relevant clusters (see Table \ref{ResultsTable}). In terms of \emph{relative} error on the enclosed mass, compared to the LTM reference set of models, we find that the median relative error on enclosed masses, $M_{2D}(<\theta\simeq136\arcsec$), is $\sim28\%$. To examine how much the squareness-artifact (\S \ref{mukappaSysSec}) may affect the discrepancies, we also examine the same difference as above inside the ``artifact-free'' region ($\lesssim120\arcsec$), finding that the median, relative systematic difference in $M_{2D}(<\theta\simeq120\arcsec$) between the two models is somewhat smaller, with a median of $\sim25\%$, implying that most of the discrepancy originates from the different parametrization, and that the boundary artifact in the LTM model contributes only about $\sim10-20\%$ to the discrepancy, near the edges. In that respect, we recommend using the current LTM models up to 2 arcminutes in radius (although we checked that concentrating only on the inner $[4\arcmin \times4\arcmin]$ does not change the statistical results, or differences between the two methods, obtained here from the full $[4.6\arcmin \times4.6\arcmin]$ FOV).

Fig. \ref{tryprof1} also reveals that the LTM method generally yields a shallower outer mass profile (and thus a higher enclosed mass), than the PIEMDeNFW model. In Fig. \ref{StackedKapProf}, we plot the stacked (i.e. averaged in radial physical bins over the 23 clusters) mass density profile from both methods. Despite evident disagreement in the mass profiles of each individual cluster (e.g. Fig. \ref{tryprof1}), between the two methods, the two \emph{stacked} profiles from the two methods usually agree within the 68.3\% confidence intervals, deduced by the scatter in each bin of the 23 profiles. For the PIEMDeNFW model we measure a decline in surface mass density ($\Sigma$, in [$g/cm^{2}$]) with physical radius, r [kpc], of $d\log (\Sigma)/d\log(r)$=-0.71, in the radial range [5,350] kpc, and for the LTM model we measure a decline of $d\log (\Sigma)/d\log(r)$=-0.57 in the same range, in excellent agreement with previous LTM analyses of well studied clusters \citep[e.g.][]{Broadhurst2005a,Zitrin2009_cl0024}. For the combined sample from the two methods together, we correspondingly obtain a slope of $d\log (\Sigma)/d\log(r)$=-0.64 in that range, whereas the errors on these slopes are roughly $\pm0.1$. 

\subsubsection{Statistical Uncertainties: Call for Caution}\label{statistical}
The statistical uncertainties are naturally coupled to the $\sigma$ errors plugged into the $\chi^{2}$:  $\sigma_{pos}$, the positional uncertainty in multiple images' location, and $\sigma_{ell}$, the WL shape measurement uncertainty. Smaller $\sigma$ values will generally entail smaller statistical uncertainties. Here, we adopted sigma values following recent works, most notably that of \citet{Newman2013} who investigated which SL positional uncertainty is preferable to consistently combine the SL constraints with WL shape measurements, whose error is generally well known ($\sigma_{ell}\sim0.3$). They found that a value of $\sigma_{pos}=0.5\arcsec$ works best, e.g. with respect to the Bayesian evidence as a criterion. This value is indeed often used in SL analyses.

As \citet{Newman2013} also mention, however, this value of 0.5\arcsec does not account for the contribution of foreground or background structure (e.g. Large-Scale Structure; LSS) along the line of sight, or other complex substructures in/near the cluster itself which may have been disregarded in the modeling. Similarly, in our previous SL analyses, we therefore usually used a SL sigma value of $\sigma_{pos}=1.4\arcsec$, which we have found takes into account modeling uncertainties arising from e.g. contribution of LSS \citep[e.g.][]{Jullo2010,D'AloisioNatarajan2011LOS,Host2012LOS}. This cosmic noise has a noticeable impact on deep SL observations in the cluster core, where magnified sources lie at greater distances \citep{Umetsu2011b}, so that it has to be taken into account especially when deep HST observations are combined with e.g. shallower ground-based WL observations. This conservative error of $\sigma_{pos}=1.4\arcsec$ was indeed found to be more realistic when the SL mass profile is combined with outer WL measurements from Subaru observations \citep{Umetsu2012}.  When accounting for possible systematic uncertainties due to prior assumptions inherent to SL modeling, we found even larger errors for $M_{2D}(<\theta)$ from SL (see Section 4.3 of \citealt{Umetsu2012} for their regularization technique), which resulted in a $\sim 20\%$ uncertainty on the total projected mass enclosed within effective Einstein radius. This is consistent with our representative systematic uncertainty on the Einstein mass estimate ($\sim$15\%).

We therefore conclude that statistical errors arising from a choice of $\sigma_{pos}$ of 0.5\arcsec~ are likely much underestimated (i.e., they neglect the governing systematic errors). We therefore apply here nominal, minimum errors on various quantities. For example, we \emph{a-priori} adopted throughout $10\%$ and $15\%$ nominal errors on the Einstein radius and mass, based on our previous analyses \citep[e.g.][]{Zitrin2012CLASH1206}, overriding the ``official'' statistical 1-$\sigma$ errors by an order of magnitude. Not surprisingly these are also similar to the $\emph{systematic}$ errors we find here for these quantities, between the two modeling methods. These minimal errors are important not to underestimate the true errors, and also since our models and $\emph{statistical}$ error maps are made publicly available online and may be used in future studies. While it may be relatively easy to rescale errors resulting from a 1-term $\chi^{2}$ to any desired $\sigma$ value, it may not be trivial for a two-term $\chi^{2}$ such as for a SL+WL combined analysis. For this reason, we chose one cluster and reran our complete analysis with $\sigma_{pos}=1.4\arcsec$. While this somewhat under-weights the SL constraints relative to the WL constraints, it will teach us by how much the typical errors increase. We find, that errors on the integrated mass profile are $\sim40\%$ larger per bin, on average, when using the more realistic $\sigma_{pos}=1.4\arcsec$ than the errors when using $\sigma_{pos}=0.5\arcsec$.

\begin{figure*}
\centering
\includegraphics[width=110mm,trim=0mm 0mm 0mm 0mm,clip]{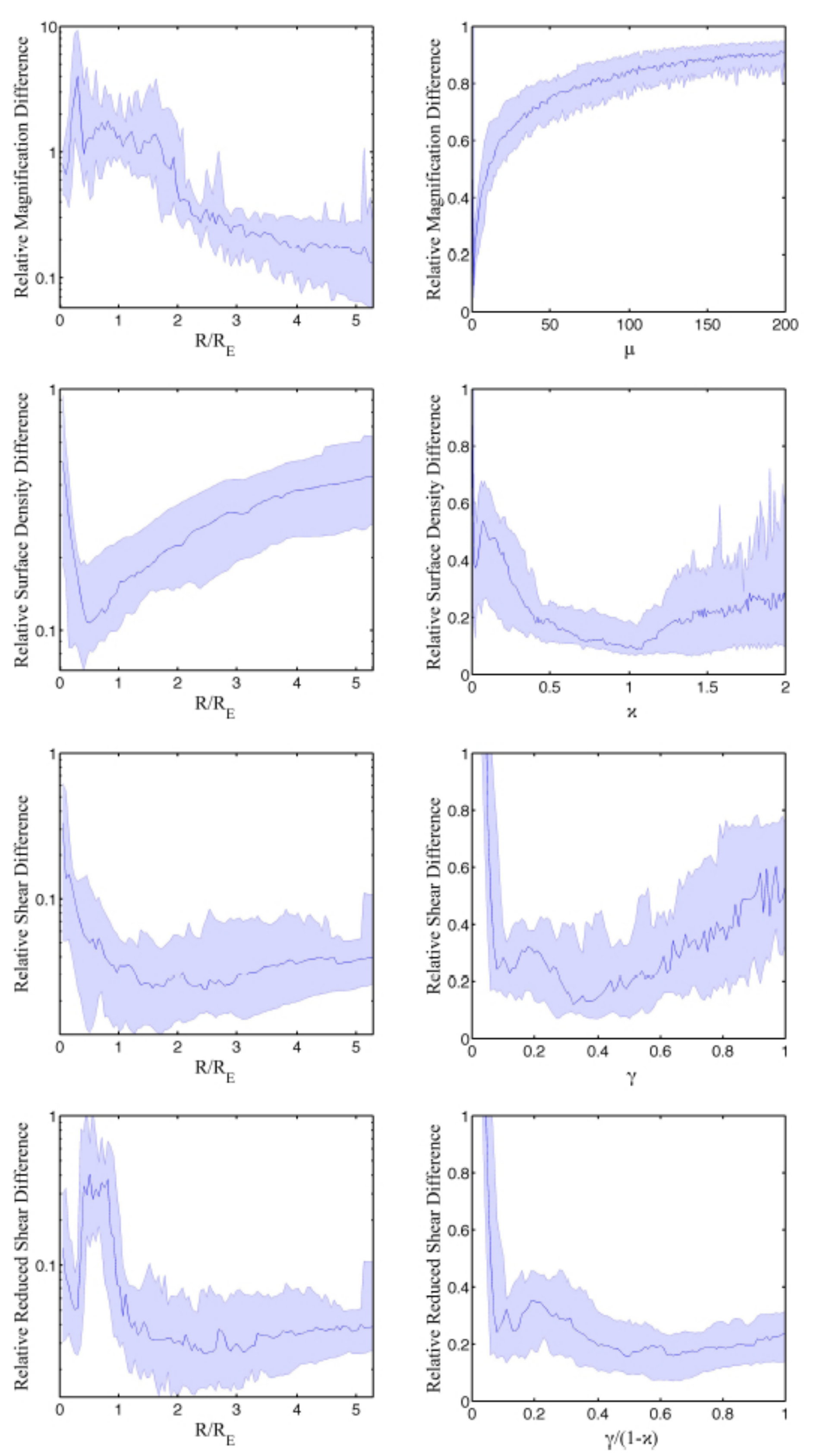}
\caption{Systematic differences, relative to the LTM model, of the magnification (\emph{top row}), surface density (\emph{second row}), shear (\emph{third row}), and reduced shear (\emph{bottom row}), as a function of the radius from the center in units of Einstein radius (\emph{left}), and as a function of the respective best-fit values of these quantities (\emph{right}). The plots are obtained by (median-) stacking the 23 clusters that have models in both parameterizations, and the shaded area represents the 1$\sigma$ confidence limit (following the scatter in each bin). The \emph{top row} shows that the radially averaged, systematic magnification difference decreases with radius from the center, and that this difference increases rapidly with magnification value so that larger magnifications have larger relative errors. The \emph{second row} shows that the radially averaged surface density difference, as expected, is minimal at about half the Einstein radius, where kappa is close to unity. The \emph{third row} shows that the mean difference in the bin-averaged shear as a function of radius is roughly constant throughout most of the range, and is significantly smaller than the error on kappa (although it can be higher for shear values close to zero or one). This is an important point: the major factor causing differences between the two models is the overall ellipticity that is being in one case assigned directly to the mass distribution and in the other case implemented as an external shear not affecting the mass distribution shape. This may create a prominent difference in the kappa maps, yet does not affect the shear (that can be similar whether it stems from the mass distribution ellipticity or is directly the external shear) to a distinguishable extent. The \emph{bottom row} shows, for completeness, the radially averaged differences in the \emph{reduced} shear. Here the behavior is similar to that of the shear, with a ``bump'' where kappa is roughly unity, boosting the reduced shear. Overall, it is evident from these figures and from our analysis that the two parameterizations cannot be easily distinguished with the strong and weak lensing data used. Additional information e.g. on the magnification, might come in handy to break this inherent degeneracy in the origin of ellipticity.}\label{hists}
\end{figure*}

We therefore recommend that for future studies, the statistical errors arising from our present analysis (i.e. with $\sigma_{pos}=0.5\arcsec$) be replaced with the actual and much larger systematic uncertainties we find in this work, to represent more realistically the true underlying (statistical+systematic) errors. These are summarized in the Abstract and in \S \ref{summary}.

\subsection{Quality of Fit and Comparison of the Two Methods}

When modeling a sample of clusters with two distinct parameterizations, a natural question arises: is there a statistically preferable parametrization? From our analysis we cannot unambiguously, strongly prefer one parametrization over the other, and the current study mainly sharpens the differences between them, and thus, advantages and disadvantages of each method (for previous examination of the differences between these parametrization, see e.g. \citealt{Zitrin2013M0416}).

Statistically, the PIEMDeNFW model seems to yield usually a more $\emph{accurate}$ and well-behaved result. This is reflected for example in the $rms$, reduced $\chi^{2}$, or Bayesian evidence, which are often (in 16 out of the 23 clusters modeled analyzed with the two methods) better for the PIEMDeNFW parameterizations, suggesting it is statistically preferable in most cases. For example, the natural logarithm of the Bayesian evidence (Table \ref{ResultsTable}) is on average, typically larger by a few dozens, for the PIEMDeNFW models in these 16 cases for which the fit is better than that of the LTM. However, recall that in \S \ref{statistical} we emphasized that the statistical errors here are strongly underestimated because of the choice of $\sigma$'s for the $\chi^{2}$ terms. Replacing for example the positional uncertainty with the more realistic value of 1.4\arcsec, the Bayesian factor comparing the two methods should become less significant -- typically the expected difference will be a factor of $\sim8$ smaller -- though still, mildly preferring the PIEMDeNFW model.

\begin{figure}
\centering
  \includegraphics[width=95mm,trim=100mm 0mm 0mm 0mm,clip]{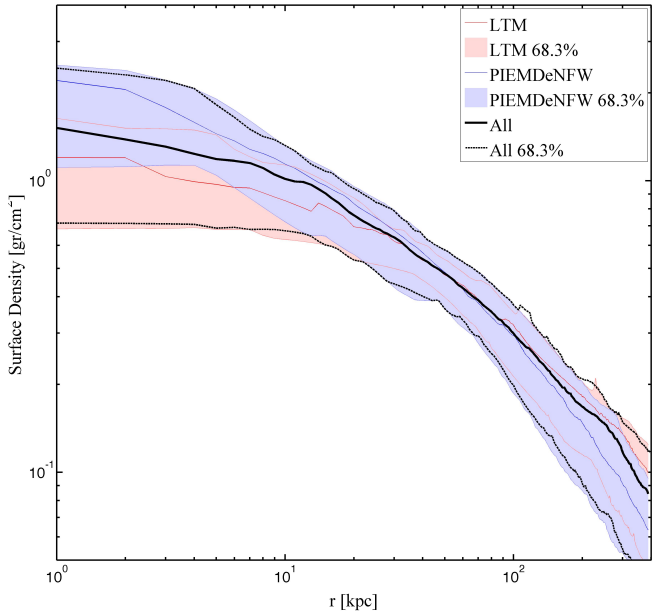}
  \caption{Stacked mass-density profile. The plot shows the projected, radially averaged mass density in $g/cm^{2}$, as a function of radius from the center in physical units (kpc), averaged over the 23 clusters that were modeled with both parameterizations. The \emph{red} plot shows the stacked LTM profile, the \emph{blue} shows the stacked PIEMDeNFW profile, and the \emph{black} lines represent the combined stacked profile and 68.3\% confidence intervals. As can be seen, the LTM profile is systematically shallower than the PIEMDeNFW profile. For more details see \S \ref{massprofiles}.}\label{StackedKapProf}
  \end{figure}

Additionally, note that if the PIEMDeNFW model is often, somewhat more $\emph{accurate}$ in term of $rms$, we regard the LTM parametrization as often more $\emph{reliable}$ (at least as a first-guess simple solution), since it relies on a very simple assumption entailing a remarkable prediction power to find many multiple images even when the fit is obtained by ``just following the light'' without any initial multiple images constraint as input \citep[e.g.][]{Broadhurst2005a,Zitrin2009_cl0024,Zitrin2012UniversalRE}, and since it is not coupled to a certain analytic form and thus allows for a more flexible profile shape (it is only coupled to the light distribution). Recall that on the far end of the ``accuracy'' scale lies non-parametric modeling \citep[e.g.][]{Abdelsalam1998,Diego2005Nonparam,Liesenborgs2006,Coe2008LensPerfect} in which the solution is $\sim$perfect in terms of multiple-image location reproduction $rms$. In such methods, typically, no prior assumptions are applied to the mass distribution and the result is directly inferred from the set of constraints; but the typical low number of constraints relative to the FOV or grid size results usually in a very low-resolution solution, with hardly any predictive power to find additional multiple images. So it is clear that  \emph{accuracy} does not necessarily mean \emph{reliability}.  Additionally, besides its immense prediction power, the fact that the same simple LTM procedure reasonably fits any lens from galaxy-scale lenses, through galaxy groups and relaxed clusters, and up to highly complex clusters, without a need to add additional DM clumps such as in the PIEMDeNFW parametrization, adds even more to its assumption reliability. On the other hand, as the LTM mass distribution is strongly coupled to the light distribution, bright galaxies that are not necessarily as massive as their luminosity indicates (or vice versa), meaning galaxies that deviate strongly from the effective $M/L$ ratio adopted, can strongly affect or bias the result. In such cases a good eye for lensing and a user intervention is needed, more significantly than in the analytic PIEMDeNFW. Such bright galaxies, if included, can artificially boost the critical curves, which may have led to an overestimation of the Einstein radius sizes in previous works (e.g. for 2-3 clusters in \citet{Zitrin2011_12macsclusters}, but based on poorer HST data - prior to CLASH imaging).

To summarize, it seems that in most cases the analytic, PIEMDeNFW model supplies more accurate results, and therefore is likely preferable for ``final'', precise lens models, while the LTM advantages are its simplicity and initial prediction power that can be used, for example, to find many sets of multiple images for new clusters. We conclude, therefore, that both are equally valuable and useful, each for its own advantages, which is why it was interesting testing the systematic differences between them, and why it is important modeling clusters with more than one method (such as in the HFF program) for a better grasp of the systematics.

\subsection{Mass-Sheet Degeneracy?}

The Mass-Sheet Degeneracy - or more generally - Transformation (MST hereafter), is a fundamental degeneracy in lens modeling \citep{Falco1985MST,Liesenborgs2012MST,SchneiderSluse2013}. Since we use parametric modeling, coupled with usually at least two meaured/fixed redshifts for multiple objects in each cluster, the MST is expected to be readily broken and thus the differences that we see between the models should not be attributed to MST (also, MST does not alter the isocontour shape or of the mass distributions, which as we saw, is the main cause of difference here originating from the difference in ellipticity assignment).  It is, however, interesting to test this assumption. For that purpose, for each field we calculated (in the ``artifact-free'' area, $\theta<120\arcsec$) the value of $\lambda$, the constant in the MST, given by: $\lambda=\sum_{i}(\kappa_{i,LTM}(\kappa_{i,NFW}-1))/\sum_{i}(\kappa_{i,LTM}(\kappa_{i,LTM}-1))$. The MST would then be $\kappa_{LTM}\rightarrow\kappa_{LTM}\lambda+(1-\lambda)$, and $\gamma_{LTM}\rightarrow\gamma_{LTM}\lambda $ (where $\gamma_{LTM}$ refers to the intrinsic shear in the LTM models, i.e. neglecting the external shear). We repeated this calculation and transformation for each LTM maps of the 23 clusters that were analyzed with the two methods, and repeated the tests for systematics differences we described in this work, now between the PIEMDeNFW maps and the MST-corrected LTM maps. If the MST really accounts for the differences between the two mass models, the differences should vanish. From this investigation, we reach the following conclusions:

$\bullet$ Accounting for MST reduced the typical relative differences in $\kappa$ maps from $\sim40\%$ to $\sim30\%$ - so that major differences still remain. MST therefore cannot account for the differences between the mass maps.

$\bullet$ If the MST could account for the differences between the mass maps, we should see a (at least roughly) constant $\lambda$ value across the FOV for each cluster. In contrast, we get that in each cluster $\lambda$ changes significantly accross the FOV: while we get typical $\lambda$ values of 1-1.3, the standard deviation across the FOV is of order $\sim0.4$ - again showing that MST is not the main reason for the differences between the methods.

$\bullet$ Examining the individual MST-corrected maps, compared to the original LTM maps, we see that the isodensity contours - and in that sense also the effective ellipticity, as well as the critical curves, remain identical to the original maps (as expected, see \citealt{SchneiderSluse2013}). This once more shows that the difference between the maps cannot be attributed to MST. Instead, as can be seen immediately from Figs. 14-16, the main difference arises from a different ellipticity of the mass distributions.

$\bullet$ In addition, the MST-corrected maps often exhibit unrealistic properties such as $\kappa<0$ at as near as few $\theta_{e}$ from the center - where this cannot be reasonable (e.g. we know from independent, larger-scale lensing analyses that this is definitely not the case). These unrealistic values indeed (mathematically) help reduce the systematic differences between the integrated mass \emph{profiles} (to $\lesssim10\%$) and average magnifications (can improve the differences by a factor of $\sim2$) - but these are physically unreasonable corrections and as was shown above - meaningless here.

We therefore conclude that MST cannot account for the differences between the mass models from  the two methods employed here. We remain with the conclusion that the differences between the models apparently arise, mainly, from the different ellipticity of the mass maps. This could have been expected, as the MST correction does not change the shape (i.e. ellipticity) or isodensity contours of the input map \citep{SchneiderSluse2013}. In that sense it is worth noting also that the inclusion of an external shear - or the ellipticity degeneracy - is a different degeneracy than the MST. If the ellipticity degeneracy were a particular case of the MST, we should have found that the external shear should be equal to a constant times the intrinsic shear, across the FOV, i.e. it could be described as $\gamma_{ext.}=k\gamma$, where $k$ is a constant. This is clearly not the case here: $\gamma$ has a typical standard deviation of $\gtrsim0.1$ from several maps we checked by eye. We conclude that the ellipticity  degeneracy therefore controls the systematic differences here, and seems to be a prominent systematic uncertainty in SL+WL analyses of galaxy clusters more generally, as shown in this work.

\subsection{A Note on Online Availability and Future Work}\label{future}
As we specified, both a comparison of our lensing 2D integrated mass profiles (Fig. \ref{tryprof1}) and the Einstein radius distributions, respectively, will be used in two upcoming works to examine their consistency with wide-field WL analyses (Umetsu et al., in preparation) and with numerical simulations (Meneghetti et al, in preparation). The comparison to independent mass profiles from wide-field WL data will both help test which of the two models agrees better with the larger-field WL data, and, will include also an overall fit to the lens models to establish e.g. the concentration-mass relation. The comparison with numerical simulations, will both examine the sample lensing and mass properties in comparison to $\Lambda$CDM to check for consistency, and, can help shed light on the underlying ellipticities of the matter distributions of CLASH-like clusters. Such comparisons will also be interesting to test for agreement with halo virilization times from simulations, or baryonic versus DM content and shape.

In that sense, it is important to mention that ellipticity could also, in principle, be added in the LTM parametrization directly into the mass distribution, rather than as an external shear. For example, in \citet{Zitrin2013M0416} we analyzed the CLASH cluster M0416 with the LTM parametrization, but with no external shear. Instead, ellipticity was embedded directly into the DM distribution by smoothing the galaxy component with an elliptical Gaussian kernel (instead of a circular one, see \S \ref{LTM}). If the underlying ellipticities of CLASH-like cluster, for example, are found eventually to be more elongated than can be described by our present LTM analysis, then it would be worth exploring further such alternative prescriptions.

\begin{figure}
\centering
  \includegraphics[width=95mm,trim=20mm 10mm 5mm 10mm,clip]{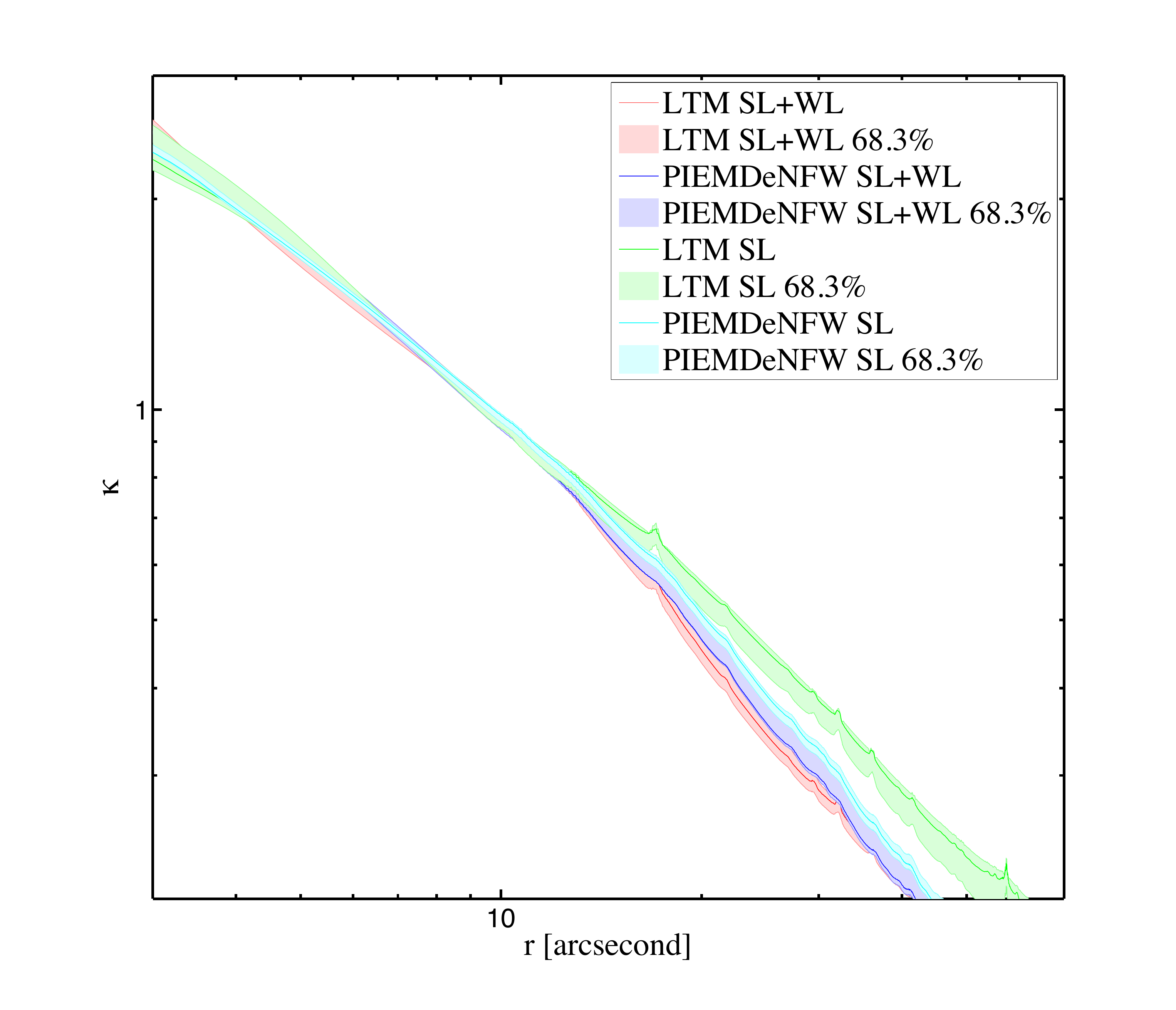}
  \caption{Effect of including the WL data, in addition to the SL data, on the resulting mass-density profile of one example cluster (MS2137), in both parameterizations. While the inclusion of HST WL here only mildly affects the PIEMDeNFW profile compared to its SL-only profile, it significantly affects the more free-form, LTM profile, improving it by about a factor of two in the outer radii (see \S \ref{future}).}\label{CompareMS2137}
  \end{figure}

It is worthwhile to mention that any lensing analysis generally calls for a possible future improvement. With time, clusters are likely to gain more exposures with HST, more multiple-images can be uncovered, and spectroscopy may be obtained for multiple images that lack accurate redshifts. All these new data will of course help to refine the lens models even further. But even prior to that, any lens model is also user-dependent and can be practically always (even slightly) improved when probing a larger and more refined parameter space. Given the volume of this work, i.e. analyzing 25 clusters with two different techniques while finding in many of them new multiple images, it is reasonable to assume that -- in contrast maybe to a work devoted to one single cluster -- there is room for future refinements of the models. Also, because we compare two methods, we tried as much as possible to minimize the user intervention (e.g. a few refinement iterations of the models are generally needed, following our past experience) so that models are, roughly, a direct and ``nearly-blind'' result of the two analysis pipelines for a similar set of input constraints. Recently we have uploaded to the MAST archive lens models for all CLASH clusters. These have been later revised, and the newer versions are those included in this work. The models being included here are being uploaded online to the MAST archive as our V2 (HST, SL+WL), CLASH lens models for the community. It is possible that newer versions will be supplied in the future in the same format presented here, and potential users will be thus referred to this work for details.

Lastly, we would like to emphasize again (see \S \ref{statistical}) that the \emph{statistical} errors here were optimized to account simultaneously for the SL and WL signals. They therefore do not contain an account for other realistic sources of error, such as contribution from foreground and background LSS etc. We advise for those directly using our models available online to adopt as nominal errors the $\emph{systematic}$ errors we found in this work, for a more responsible error budget.

On the same matter, it would be very useful to check in a future study, how much the WL data actually add to (or affect) the overall fit, which is particularly interesting to examine in the HST WL regime (i.e. beyond the SL regime, well outside the Einstein radius) to see if the WL constraints refine the outer mass profile, for example. Due to the extent of this work we do not attempt to thoroughly pursue that study here, which would require remodeling the full sample with only SL data for comparison, but as a preliminary, general test we adopted one cluster from our list (MS2137) and remodeled it using both parametrizations, now without the WL input. The resulting mass profiles are seen in Fig. \ref{CompareMS2137}. We find that the PIEMDeNFW model is not significantly altered by omitting the WL data, but the (more free-form profile shape) LTM mass profile did significantly change beyond the SL regime. The LTM model that did include the WL measurements is much closer to the PIEMDeNFW profiles, than the LTM model that did not include the WL data. The mass-density profile of the LTM model with no WL data differs on average by $22\%$ from the PIEMDeNFW mass-density profile, and by $12\%$ from the LTM model that included also the WL data, in the radial range 35\arcsec-120\arcsec (35\arcsec corresponds to twice the Einstein radius), whereas the LTM model that included also the WL data only differs by $13\%$ from the PIEMDeNFW kappa profile in the same radial range. This shows that at least for LTM model, whose profile is more free than that of the PIEMDeNFW model since it is not coupled to a certain analytical form, the WL data do help refine the fit and pin down the mass profile, as could be expected. The improvement on the outer mass profile, compared to the PIEMDeNFW model (if referring to the latter as a reference), is nearly a factor of two. As a second, rapid test, we also constructed one model for this cluster using only the WL data, with the PIEMDeNFW parametrization. Although we do not show it explicitly here, we note that the resulting model has a similar mass profile throughout, where only the normalization is missing (as expected from the mass sheet degeneracy, broken when e.g. adding SL constraints). Lastly, we used the two models for MS2137 constructed using SL constraints only, without WL information, and tested the agreement with the WL data for these. The PIEMDeNFW model slightly, better agrees with the WL data, than the LTM model: the WL $\chi^{2}$ for the PIEMDeNFW SL-only model is $\simeq1545$, whereas the WL $\chi^{2}$ of the LTM SL-only model is only $\sim2\%$ higher, $\simeq1580$. For comparison, the WL $\chi^{2}$ for the SL+WL models is $\simeq1540$ for both parameterizations\footnotemark[2] \footnotetext[2]{the value differs than that in Table \ref{ResultsTable} since the test here was performed with a different, lower resolution.}. Assessing the effect of HST WL data on the overall fit for the statistical sample will also be very important for the ongoing HFF campaign, for which future versions of refined lens models will be constructed, possibly using both HST SL+WL deep data, in order to secure the magnification predictions beyond the SL regime. We hope to examine this more thoroughly using the full sample, in a future related study.

\section{Summary}\label{summary}
One of the main goals of the CLASH multi-cycle treasury program, set a few years ago to observe 25 mainly X-ray selected clusters, has been to study their mass distributions and related properties, and confront these results with expectations for mass assembly or structure formation from $\Lambda$CDM. The CLASH program has contributed significantly to the cluster, lensing, and supernova fields \citep[e.g.][]{Graur2014SNrate,Patel2014SN,Coe2012A2261,Monna2014RXC2248,Medezinski2013M0717,Umetsu2012,Zitrin2011CLASH383,Zitrin2012CLASH0329,Zitrin2012CLASH1206,Zitrin2013M0416}, and has uncovered, through lensing, hundreds of high redshift galaxies \citep[e.g.][]{Bradley2013highz} including some of the highest-redshift galaxies known to date \citep{Bouwens2012highzInCLASH,Zheng2012NaturZ,Coe2012highz}. Most recently, \citet{Merten2014CLASHcM} have produced the most up-to-date c-M relation derived from the CLASH sample, using non-parametric SL+WL analysis, then compared to $\Lambda$CDM simulations \citep{Meneghetti2014CLASHsim}, and \citet{Umetsu2014CLASH_WL} and \citet{Donahue2014CLASHX} have studied the WL and X-ray mass proxies and properties of the CLASH sample.

Aside from the treasury HST observations, CLASH has also been graciously granted with, or used existing, other space observations from XMM-Newton, Chandra, and Spitzer, for X-ray and IR studies; ground-based wide-field imaging from Subaru used for wide-field WL analyses; and a dedicated VLT campaign to obtain spectroscopic redshifts for the multiple-images, some of which we have presented and used in this work. Additional LBT and Keck observing times kindly granted to us in various frameworks have also enabled arc redshifts, which will be used for future refinements of the mass models.

Here, we completed the high-resolution lensing analysis of the 25 CLASH clusters, in HST data. We incorporated both the SL features and HST WL shape measurements for the full sample. We make available the mass and magnification maps to the community, and have characterized them in this work, with an emphasis on quantifying in addition to the output statistical uncertainties, also the underlying systematics. To do so we analyzed nearly all clusters with two distinct parameterizations - one adopts light-traces-mass for both galaxies and dark matter while the other adopts an analytical, elliptical NFW form for the dark matter.

We have found that the current SL+WL data alone, cannot unambiguously distinguish between an intrinsically elliptical mass distribution, or a light-tracing-mass distribution for which the overall ellipticity is introduced only in the form of an external shear not contributing to the mass distribution. These two distinct parameterizations introduce some notable discrepancies. We found that the typical (median), relative systematic differences throughout the central $[4.6'\times4.6']$ analysis FOV, are $\sim40\%$ in the (dimensionless) mass density, $\kappa$, and $\sim20\%$ in the magnification, $\mu$. We showed maps of these differences for each cluster, as well as the mass distributions, critical curves, and 2D integrated mass profiles. The Einstein radii ($z_{s}=2$) typically agree within $10\%$ between the two models, and Einstein masses agree, typically, within $\sim15\%$. At larger radii, the total projected, 2D integrated mass profiles of the two models, within $r\sim2\arcmin$, differ by $\sim30\%$. Stacking the surface-density profiles of the sample from the two methods together, we obtain an average slope of $d\log (\Sigma)/d\log(r)\sim-0.64\pm0.1$, in the radial range [5,350] kpc.

Our publicly-available models and the errors we find here, we hope, should be most useful for future high-impact studies of lensing clusters and the objects behind them. A comparison of the sample's statistical properties, for example of the Einstein radius distribution with $\Lambda$CDM, or the agreement of our mass profile with wider-field independent WL analyses, remains for future work.

\section*{acknowledgments}
We thank the reviewer of this work for most valuable comments. AZ is grateful for useful discussions with Carrie Bridge and Drew Newman. This work is based on observations made with the NASA/ESA {\it Hubble Space Telescope}. Support for program \#12065 was provided by NASA from the Space Telescope Science Institute (STScI), which is operated by the Association of Universities for Research in
Astronomy, Inc. under NASA contract NAS~5-26555. Support for this work was provided by NASA through Hubble Fellowship grant \#HST-HF2-51334.001-A awarded by STScI. KU acknowledges support from the Ministry of Science and Technology of Taiwan through the grant MOST 103-2112-M-001-030-MY3. The research was in part carried out at the Jet Propulsion Laboratory, California Institute of Technology, under a contract with NASA.
\bibliographystyle{apj}
\bibliography{outDan2Agnese}

\begin{figure*}
\centering
  \includegraphics[width=0.99\textwidth,height=1.3\textwidth,trim=20mm 35mm 20mm 20mm,clip]{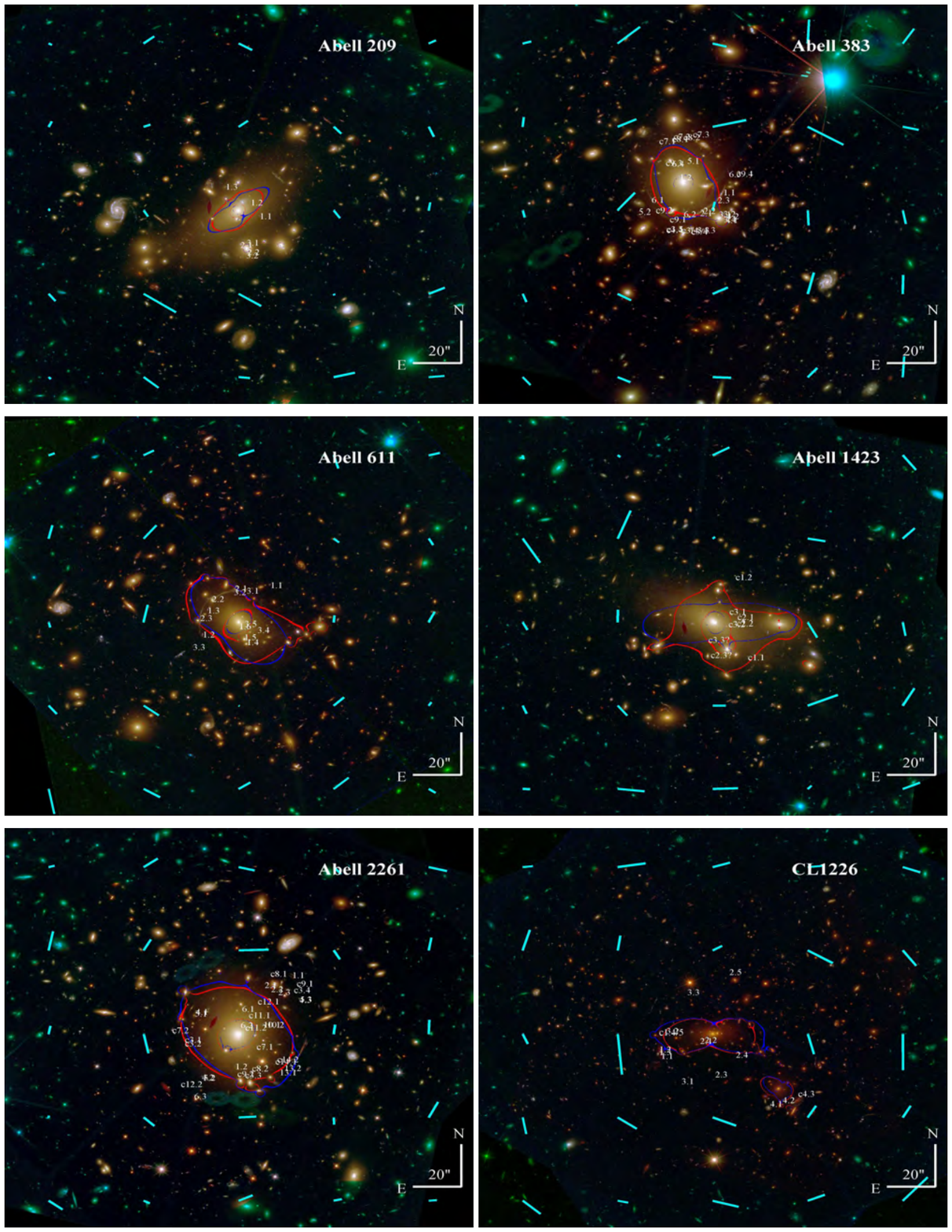}
  \caption{Multiple images and candidates, shear, and critical curves (for $z_{s}=2$), overlaid on an RGB color image constructed from the CLASH 16-band imaging, for six clusters from our sample noted on each subfigure (for other clusters see Figs. \ref{curves2}-\ref{curves4}). The \emph{red} critical curves correspond to our LTM model whereas the \emph{blue} critical curves correspond to our PIEMDeNFW model. The shear, averaged here in $\sim[40\arcsec \times 40\arcsec]$ pixels for show, is marked with \emph{cyan} lines across the field, where the line length in each position is linearly scaled with the shear's strength. Multiple images are listed in Table \ref{multTable}; the resulting mass density maps are shown in Figs. \ref{Kaps1} and \ref{Kaps2}; the resulting mass profiles are shown in Figs. \ref{tryprof1} and the differences between the various maps are shown in Figs. \ref{DifsKap1} and \ref{DifsMag1}.}\label{curves1}
  \end{figure*}

  \begin{figure*}
\centering
  \includegraphics[width=0.99\textwidth,height=1.3\textwidth,trim=20mm 35mm 20mm 20mm,clip]{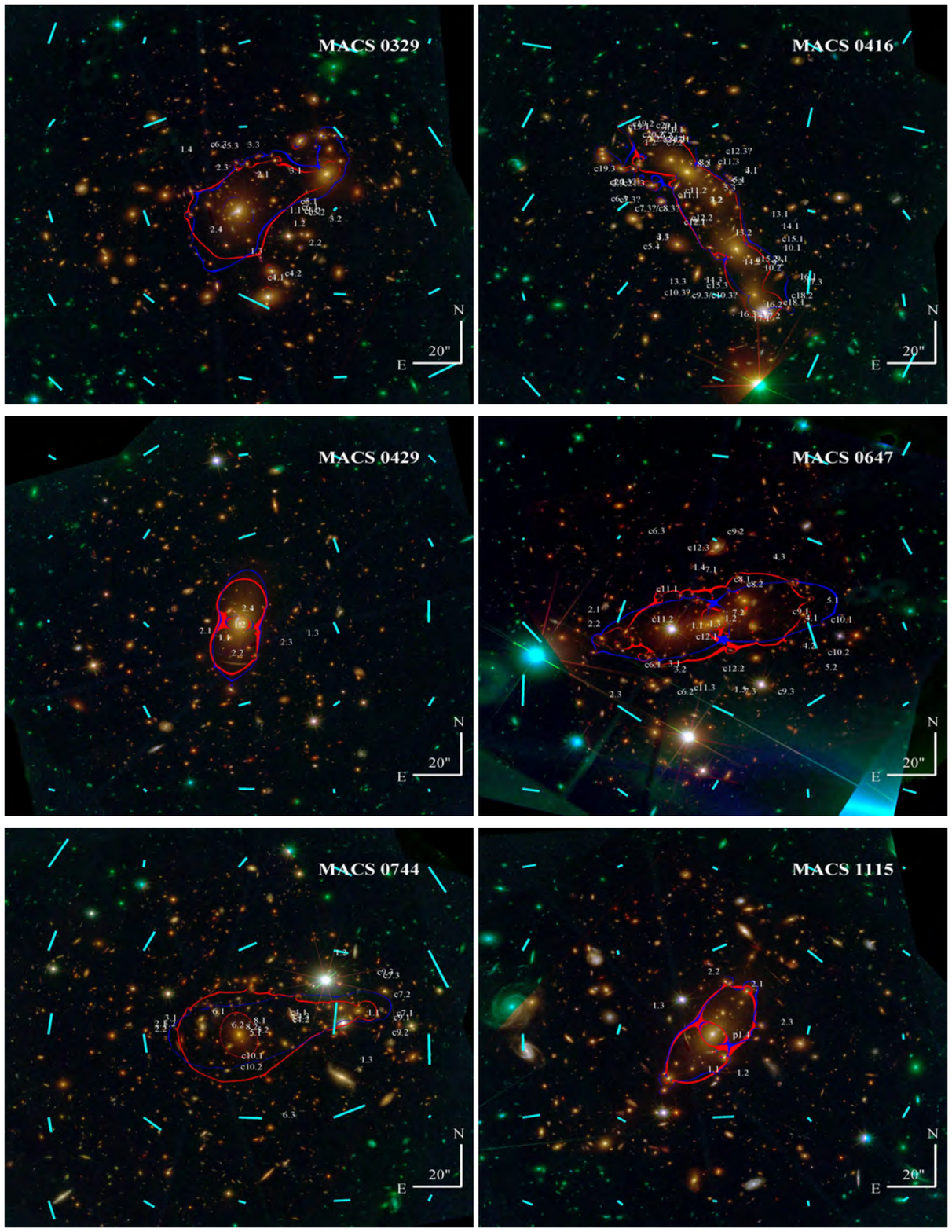}
  \caption{Same as Fig. \ref{curves1}, for another six clusters from our sample. Cluster ID's are noted on each subfigure.}\label{curves2}
  \end{figure*}

  \begin{figure*}
\centering
 \includegraphics[width=0.99\textwidth,height=1.3\textwidth,trim=20mm 35mm 20mm 20mm,clip]{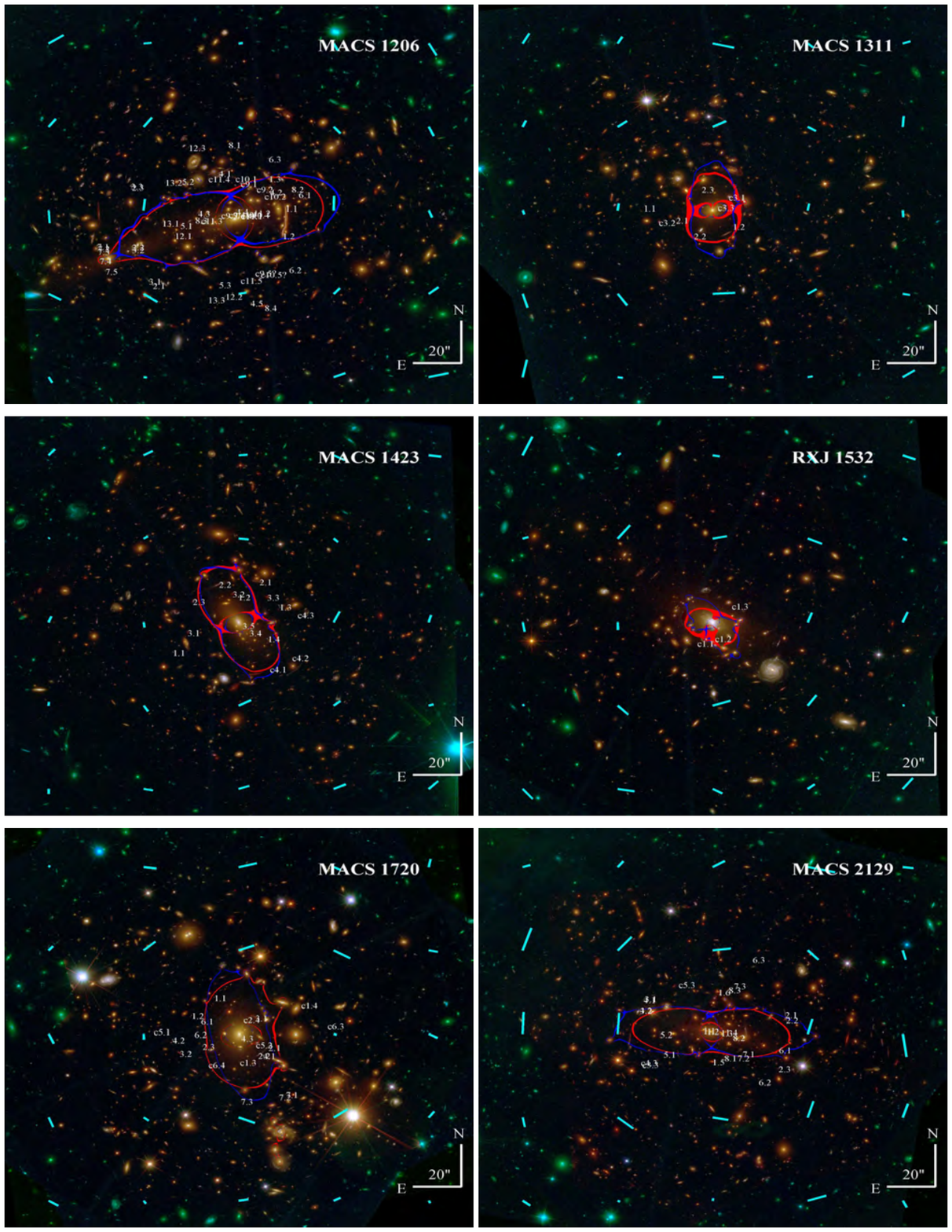}
  \caption{Same as Fig. \ref{curves1}, for another six clusters from our sample. Cluster ID's are noted on each subfigure.}\label{curves3}
  \end{figure*}

    \begin{figure*}
\centering
 \includegraphics[width=0.99\textwidth,height=1.3\textwidth,trim=20mm 35mm 20mm 20mm,clip]{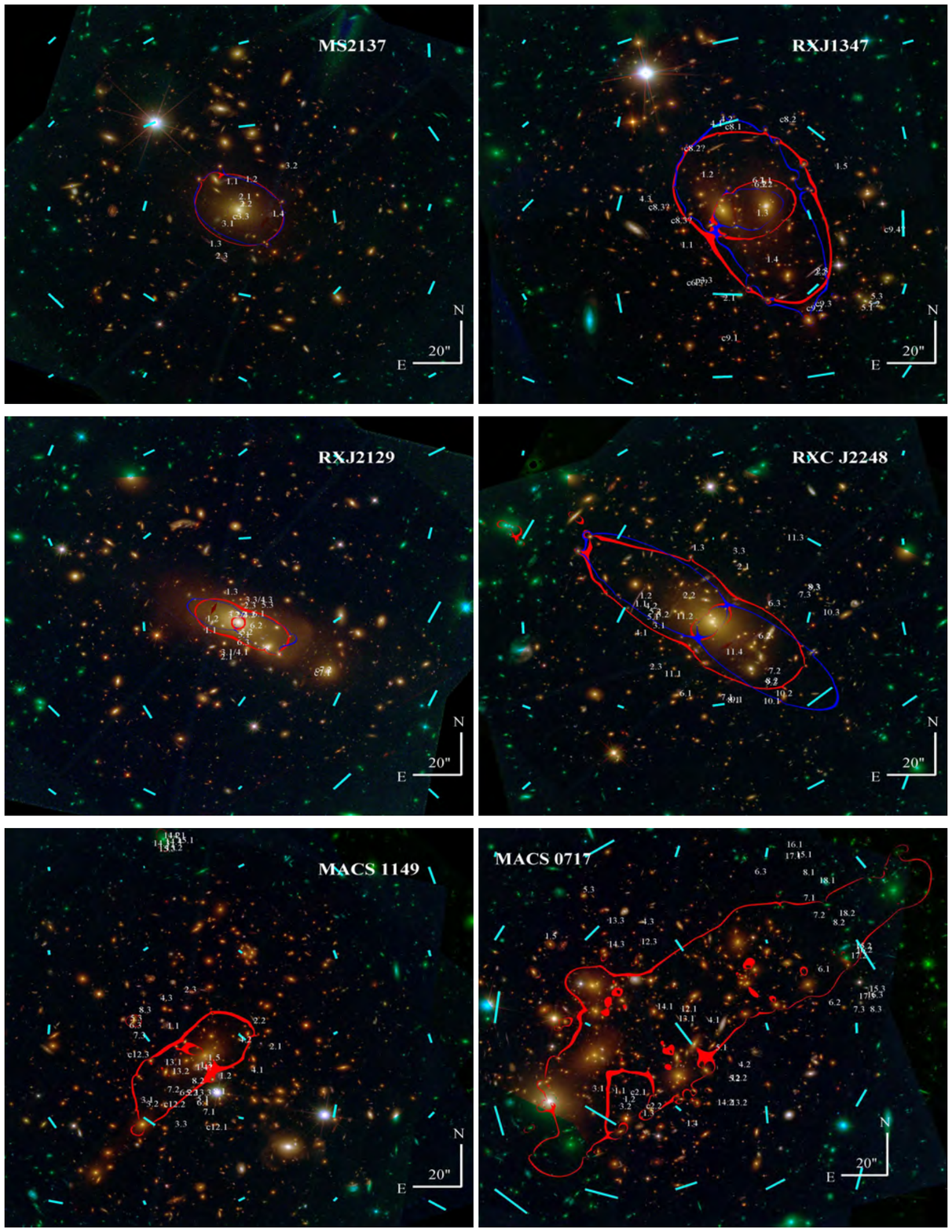}
  \caption{Same as Fig. \ref{curves1}, for another six clusters from our sample. Cluster ID's are noted on each subfigure.}\label{curves4}
  \end{figure*}

\begin{figure*}
\centering
  \includegraphics[width=0.99\textwidth,height=1.3\textwidth,clip]{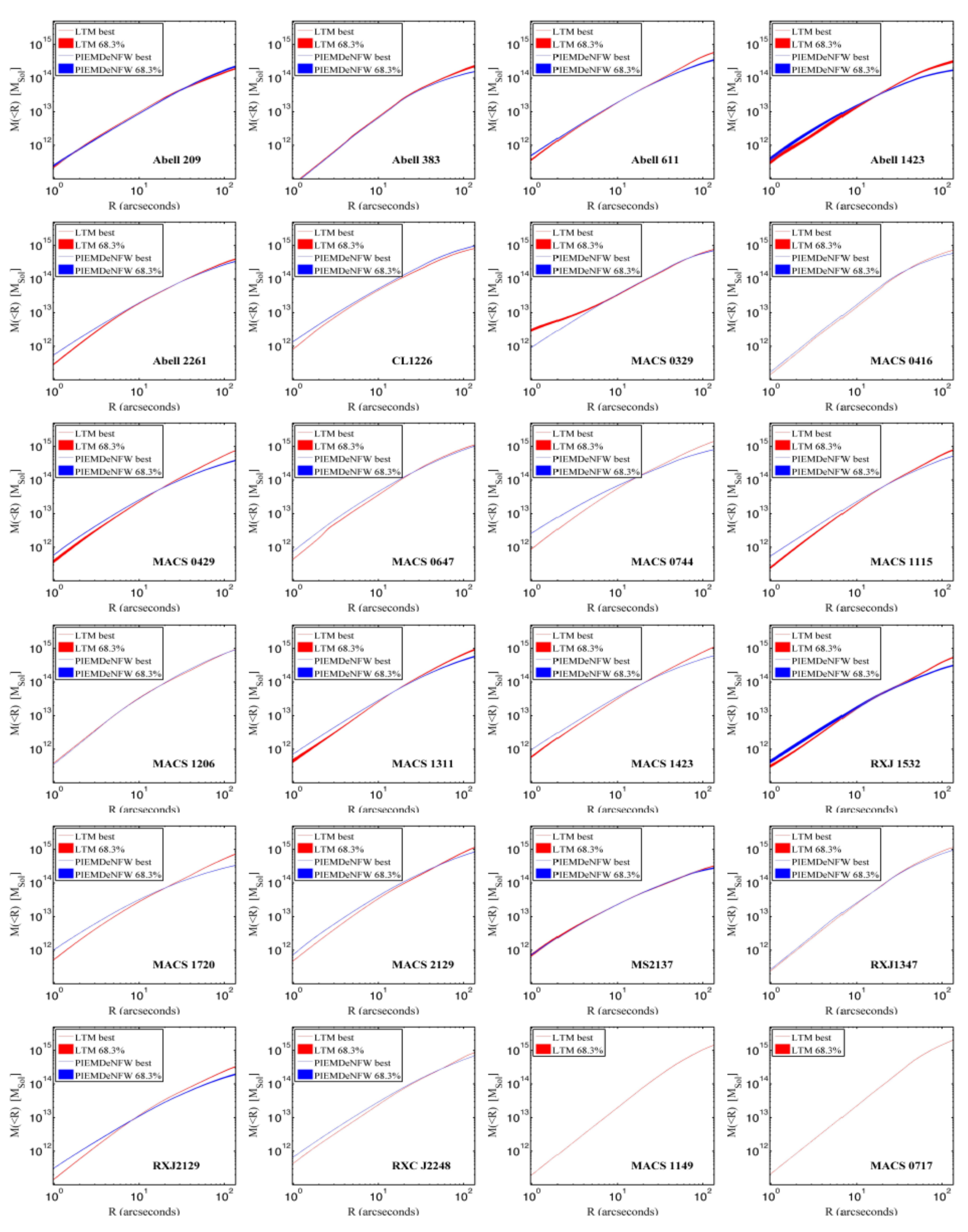}
  \caption{Integrated, 2D mass profiles for the full CLASH sample, from our HST SL+WL analysis described in this work. The \emph{red} plot shows the LTM profile and errors for each cluster, and the \emph{blue} plot shows the profile and errors of the PIEMDeNFW model. For more details see \S \ref{massprofiles}. }\label{tryprof1}
  \end{figure*}

  \begin{figure*}
\centering
  \includegraphics[width=0.99\textwidth,height=1.3\textwidth,trim=4mm 8mm 4mm 7mm,clip]{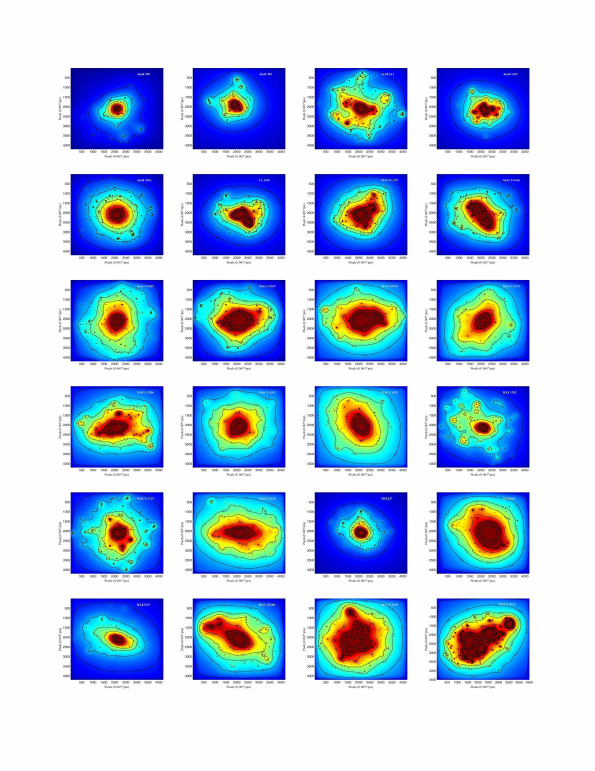}
  \caption{Stamp images showing the surface mass-density map from our LTM models, for 24 CLASH clusters noted on the stamp images (one cluster's map was shown in Fig. \ref{KapLTMNFW0} and was omitted from this composite stamp Figure). Maps are scaled to a fiducial redshift corresponding to $d_{ls}/d_{s}=1$, the adopted default for the HFF map making project.}\label{Kaps1}
  \end{figure*}

  \begin{figure*}
\centering
   \includegraphics[width=0.99\textwidth,height=1.3\textwidth,trim=4mm 8mm 4mm 7mm,clip]{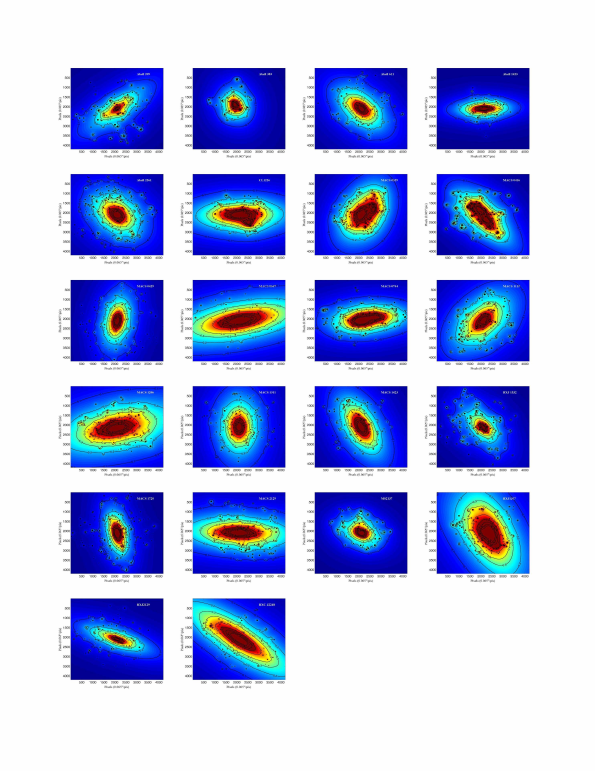}
  \caption{Same as Fig. \ref{Kaps1}, but for our PIEMDeNFW models.}\label{Kaps2}
  \end{figure*}

\begin{figure*}
\centering
  \includegraphics[width=0.99\textwidth,height=1.28\textwidth,trim=6mm 8mm 6mm 7mm,clip]{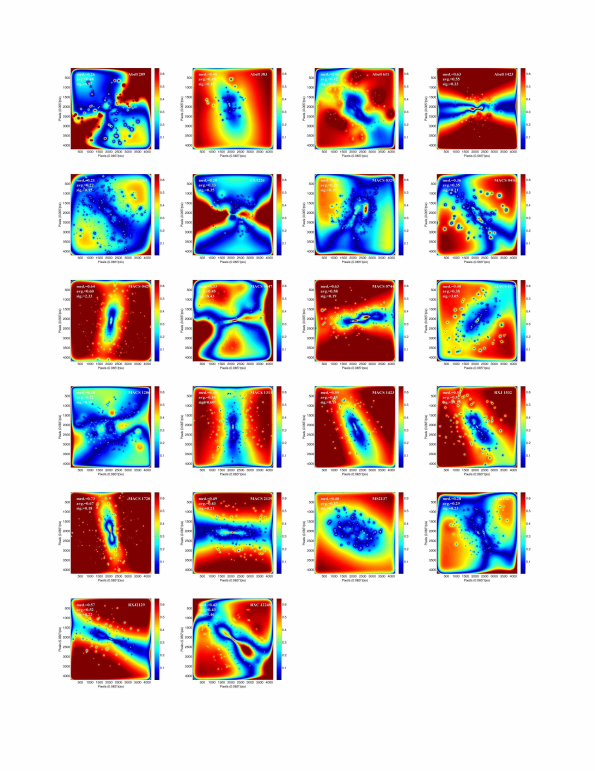}
  \caption{Absolute-value of the difference between the LTM $\kappa$ maps seen in Fig. \ref{Kaps1}, and the PIEMDeNFW $\kappa$ maps seen in Fig. \ref{Kaps2}, relative to the LTM maps which are used as references. A similar map for the 25th CLASH cluster was shown in Fig. \ref{DifsKap0}. On each subfigure we note the cluster name, average, median and standard deviation values. As can be seen, differences are mainly caused by ellipticities being assigned to the PIEMDeNFW mass density distributions directly, while the LTM mass density distributions simply follow the light and no overall ellipticity is introduced to them directly. Additionally, artifacts from the smoothing procedure introduce squareness in the LTM models near the edges of the FOV, which contributes further to the discrepancy at larger radii. We find that the typical difference in $\kappa$ throughout these FOVs is $\sim40\%$, and the distribution of differences is shown in Fig. \ref{hists}. See \S \ref{discussion} for more details.}\label{DifsKap1}
  \end{figure*}

\begin{figure*}
\centering
 \includegraphics[width=0.99\textwidth,height=1.28\textwidth,trim=6mm 8mm 6mm 7mm,clip]{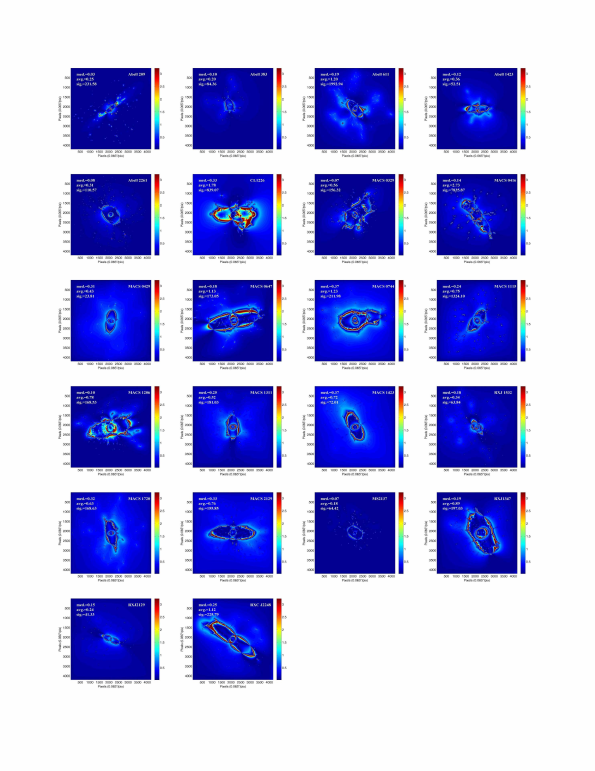}
  \caption{Same as Fig. \ref{DifsKap1}, but now showing the absolute-value differences in magnification, relative to the LTM models. A similar map for the 25th CLASH cluster was shown in Fig. \ref{DifsKap0}. As can be seen, the majority of differences are seen in the vicinity of the diverging critical curves, where farther away from them the error is much lower. We find that the typical difference in $\mu$ throughout these FOVs is $\sim20\%$, and the distribution of differences is shown in Fig. \ref{hists}. See \S \ref{discussion} for more details.}\label{DifsMag1}
  \end{figure*}

\LongTables
\clearpage
\tabmultiples
\clearpage

\end{document}